# Phase-change materials for volatile threshold resistive switching and neuronal device applications

Huandong Chen[1,4], Jayakanth Ravichandran[1,2,3*]

[1]Mork Family Department of Chemical Engineering and Materials Science, University of Southern California, Los Angeles CA, USA

[2]Ming Hsieh Department of Electrical and Computer Engineering, University of Southern California, Los Angeles CA, USA

[3]Core Center for Excellence in Nano Imaging, University of Southern California, Los Angeles CA, USA

[4]Present Address: Condensed Matter Physics and Materials Science Department, Brookhaven National Laboratory, Upton NY, USA

*Email: j.ravichandran@usc.edu

## Abstract

Volatile threshold resistive switching and neuronal oscillations in phase-change materials, specifically those undergoing ‘metal-to-insulator’ transitions, offer unique attributes such as fast and low-field volatile switching, tunability, and stochastic dynamics. These characteristics are particularly promising for emulating neuronal behaviors and solving complex computational problems. In this review, we summarize recent advances in the development of volatile resistive switching devices and neuronal oscillators based on three representative materials with coincident electronic and structural phase transitions, at different levels of technological readiness: the well-studied correlated oxide $VO_2$, the charge-density-wave transition metal dichalcogenide $1T$-$TaS_2$, and the emerging phase-change chalcogenide perovskite $BaTiS_3$. We discuss progresses from the perspective of materials development and device implementation. Finally, we emphasize the major challenges that must be addressed for practical applications of these phase-change materials and provide our outlook on the future research directions in this rapidly evolving field.

## 1. Introduction

Due to the recent slowdown in Moore's law and the increasing computational demands of artificial intelligence applications,[1-3] the development of novel "post-complementary metal-oxide-semiconductor (CMOS)" hardware that is both energy-efficient and capable of handling complex tasks has become highly sought after.[4,5] Neuromorphic computing, a paradigm that interconnects networks of artificial synapses and neuronal devices, can physically emulate the structure and function of the human brain, offering remarkably low power consumption and intrinsic learning capabilities.[6-10] Figure 1a illustrates a biological neuron specialized for processing and transmitting cellular signals, whereas Figure 1b shows a typical tonic firing pattern of single neuron exhibiting rhythmic spiking activity.[11] Early attempts to mimic the neuronal and synaptic behavior of brain used non-von Neumann architectures based on conventional CMOS circuits and metal-oxide memristors, enabling the demonstration of millions of programmable spiking neurons constructed from billions of transistors.[12,13] However, such systems remain far away from true brain-like operation, primarily due to its low energy efficiency and limited system complexity.

Alternatively, researchers have pursued novel materials and device architectures that leverage intriguing physical phenomena to emulate synaptic and neuronal functionalities. A range of non-volatile devices, including memristors,[14-16] phase change memory,[17,18] ferroelectric memory,[19-21] and magnetic tunnel junctions,[22-24] have been developed as synaptic components in integrated neuromorphic systems. Meanwhile, materials exhibiting volatile threshold switching mechanisms, such as metallic filament type,[25,26] thermal feedback,[27,28] ferroelectric,[29,30] and electronic phase transitions,[31-34] are often employed as neuronal devices. Among these, phase-change materials with coincident structural and electronic phase transitions are especially promising for mimicking the oscillatory and dynamic features of biological neurons. Upon varying temperatures, such materials often undergo well-defined jumps in their electrical resistivities across the transition, as illustrated in Figure 1c. Oftentimes, volatile resistive switching is electrically triggered on simple two-terminal devices when operating below the transition temperature (Figure 1d). By further integrating such phase-change devices into various oscillatory circuits (Figure 1e), self-sustained voltage oscillations (Figure 1f), or even more complex neuron-like dynamic behaviors such as tonic firing pattern (Figure 1b) can be realized. Table 1 shows a selected list of phase-change material candidates, primarily transition metal (V, Ti, Ni, Ta, etc.) oxides and chalcogenides, that are promising in realizing such electronic functionalities, despite the varying transition temperatures and mechanisms.

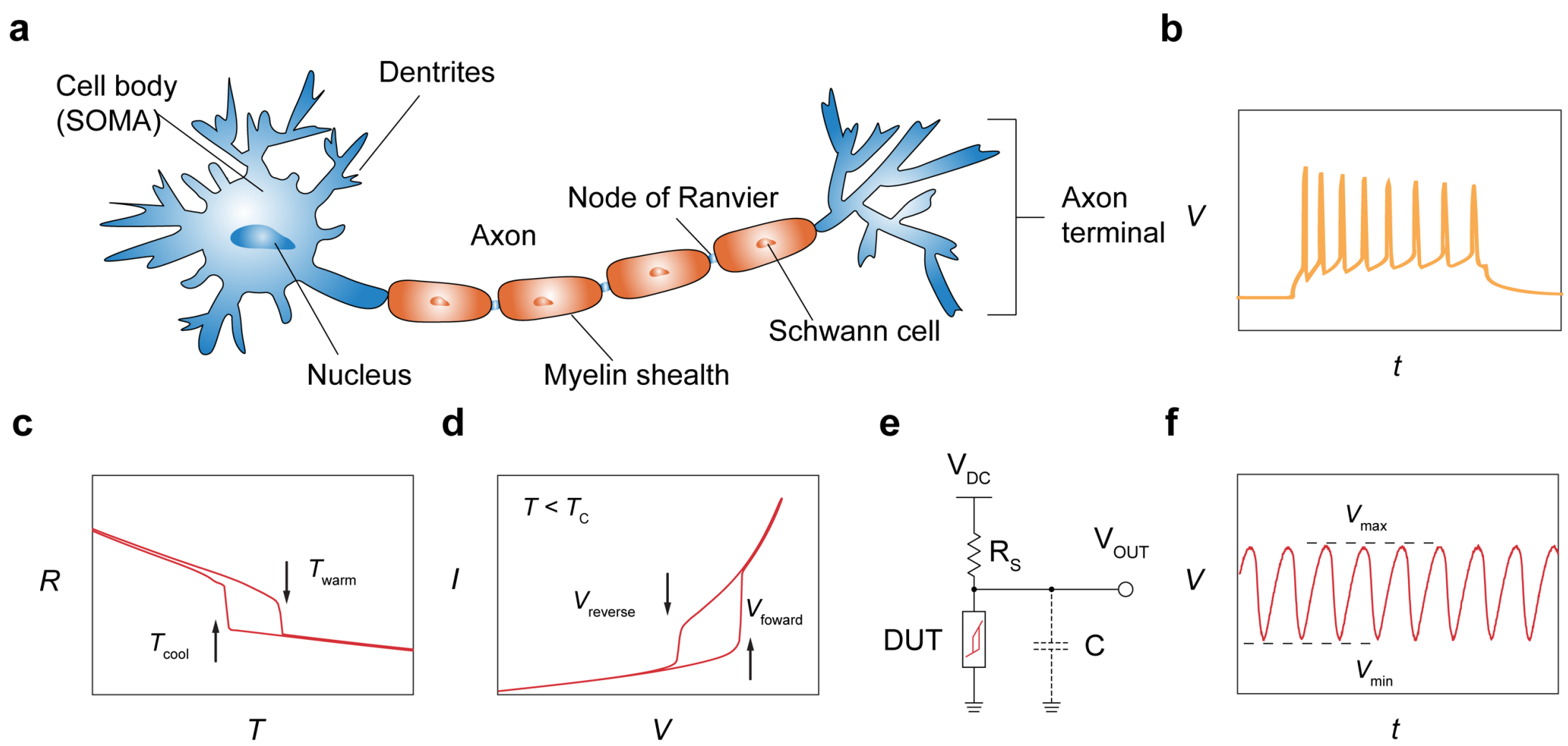

**Figure 1** (**a**) Schematic illustration of the structure of a biological neuron. (**b**) Representative tonic firing pattern of a single neuron showing spiking activity. (**c**) Typical temperature-dependence of electrical resistance of a phase-change material showing electronic phase transitions. (**d**) Typical *I-V* characteristics of volatile threshold resistive switching in a phase change material driven by DC voltages. (**e**) and (**f**) Representative circuit diagram for introducing voltage oscillations in a two-terminal phase-change volatile resistive switching device and the corresponding oscillation waveform. (**c**) to (**f**) Figures adapted from Ref. [34] Copyright 2023, John Wiley & Sons.

Notably, there have been a long history and extensive research interests on studying metal-to-insulator transitions (MIT) in various correlated binary and complex oxides. In a pioneering work in late 1950s, Morin observed that the resistance of certain binary transition metal oxides, such as vanadium oxides ($VO_2$, $V_2O_3$) and titanium sesquioxide ($Ti_2O_3$), increases by several orders of magnitude upon crossing a critical temperature ($T_C$).[35] As sister compounds of vanadium oxides, niobium oxides (e.g., $NbO_2$) also undergo similar transitions, albeit at much higher transition temperatures.[36,37] These phenomena are predominantly attributed to Mott-Hubbard transitions.[38-40] As illustrated in Figure 2a (left), the gap arises from electrons hopping between *d* bands of adjacent anions and is associated with the Coulomb interaction *U* (or Hubbard correlation energy).[38-40] Later, rare-earth nickelate perovskites, $LnNiO_3$ (where Ln = Pr, Nd, Sm), were reported to exhibit metal-to-insulator transitions, in both bulk form and strained thin films, where the gap opening is potentially originated from the excitation from the anion *p* level to the metal *d* level with the charge-transfer energy Δ, as illustrated in Figure 2a (right).[48,57] A theoretical framework describing Mott-Hubbard and charge-transfer insulators can be understood using the Zaanen-Sawatzky-Allen scheme.[40,58]

**Table 1 Selected list of phase-change material candidates for resistive switching.** Note that only the volatile switching and the associated neuronal oscillation applications are discussed in this review.

| Material | Phase Transition Properties | | | | Primary Mechanism | Device Demonstration | | Ref. |
|---|---|---|---|---|---|---|---|---|
| | $T_C$ | Structural Transition | $\rho / \rho_0$ | Year | | Resistive Switching | Electronic Devices | |
| $VO_2$ | 340 K | $P4_2/mnm$ to $P2_1/c$ | $10^2$-$10^5$ | 1959 | Electron-electron interactions (Mott-Hubbard + Charge-transfer) | Volatile | Oscillator | 35,41 |
| $V_2O_3$ | ~165 K | $R\bar{3}c$ to $I_2/a$ | $10^6$ | 1959 | | Volatile | Oscillator | 35,42,43 |
| $NbO_2$ | ~1070 K | $P4_2/mnm$ to $I4_1/a$ | 10 | 1966 | | Volatile | Oscillator | 44,45 |
| $PrNiO_3$ | 135 K | *Pbnm* to $P2_1/n$ | $10^3$ | 1991 | | No | No | 46-48 |
| $SmNiO_3$ | 403 K | *Pbnm* to $P2_1/n$ | ~10 | | | No | No | |
| $NbSe_3$ | 145 K | Displacement of Nb (III) column | ~1.2 | 1976 | Electron-phonon interactions (CDW) | No | No | 49-51 |
| | 59 K | Displacement of Nb (I) column | ~2 | | | Volatile | No | |
| 1*T*-$TaS_2$ | 350 K | NC-CDW to IC-CDW | ~2 | 1971 | | Volatile | Oscillator | 52-54 |
| | 130-230 K | IC-CDW to CDW | ~20 | | | Non-volatile | Memristor | |
| 1*T*-$TaSe_2$ | 473 K | IC-CDW to CDW | 3-4 | 1974 | | No | No | 55 |
| $BaTiS_3$ | 250 K | Hexagonal to trigonal (CDW) | ~2 | 2023 | | Volatile | Oscillator | 56 |
| | 150-190 K | Trigonal (CDW) to monoclinic | ~10 | | | No | No | |

Charge-density-wave (CDW) materials belong to another class of promising candidates to host metal-to-insulator transitions due to the gap opening from strong electron-phonon interactions.[59] According to Peierls' picture, a one-dimensional atomic chain is inherently unstable low temperatures, resulting in a spontaneous periodic lattice distortion and gap opening at the zone boundary, as illustrated in Figure 2b.[59,60] Such transition is energetically favorable as the electronic energy saved due to the new band gaps overweighs the elastic energy consumption from lattice reconstruction.[59] However, only a handful of real CDW materials manifest pronounced resistivity jumps across transitions, which is a critical ingredient for potential device applications.[61] The 1*T* phase of the transition metal dichalcogenide $TaS_2$ happens to exhibit such metal-to-insulator transitions, as first observed by Thompson *et.al*. in the early 1970s.[52] With recent advances in 2D device fabrication and testing, interest in utilizing 1*T*-$TaS_2$ for neuronal applications has grown substantially since the mid-2010s.[54,62,63] Moreover, emerging chalcogenide crystals, such as $BaTiS_3$ and $EuTe_4$, have also shown to host both CDW orders and metal-to-insulator transitions,[34,56,64,65] rendering them promising candidates to demonstrate resistive switching for emulating synaptic or neuronal functionalities. Note that due to the large thermal hysteresis, $EuTe_4$ tends to demonstrate non-volatile resistive switching and memory effects when operating at temperatures within the hysteresis window.[65]

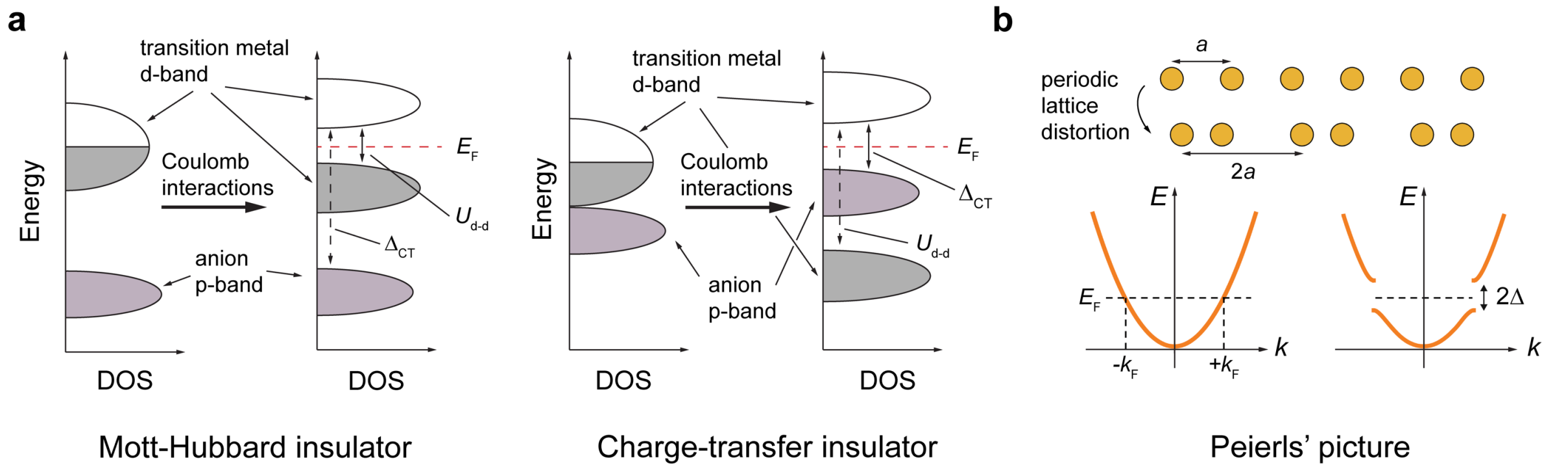

**Figure 2** (**a**) Schematic energy-band diagrams of Mott-Hubbard insulator (left) and charge-transfer insulator (right). Figure adapted from Ref. [40] Copyright 1998, American Physical Society. (**b**) Schematic illustration of Peierls instability in one-dimensional chains of atoms. Figure adapted from Ref. [60] Copyright 2021, Springer Nature.

In this work, we focus explicitly on phase-change oxides and chalcogenides that exhibit metal-to-insulator transitions and volatile resistive switching, as well as their potential for neuronal device applications. Among all the candidates that have demonstrated such functionalities, we choose three archetypal phase-change materials – namely the correlated binary oxide $VO_2$ and the CDW chalcogenides 1*T*-$TaS_2$ and $BaTiS_3$ – and discuss their similarities and differences in intrinsic structural and electrical properties, material synthesis and device fabrication methods, and electronic device implementation and characteristics. These materials represent two different categories of underlying mechanisms, i.e., electron correlation and electron-lattice interaction. Considering that there have been several comprehensive literature reviews on metal-to-insulator transitions in correlated oxides and their applications for neuromorphic computing,[66-68] only one representative oxide material, $VO_2$, was chosen from that category, as a benchmark to compare the similarities and differences with the other two phase-change compounds.

Importantly, these materials are at three very different levels of technological readiness: for instance, after being studied for more than 60 years, $VO_2$ is close to a 'technologically ready' status with well-established thin-film synthesis and device fabrication processes that are ideal for large-scale device applications; 1*T*-$TaS_2$ has been actively studied at individual-device levels in research-laboratories over the past 10-15 years, however, such mechanical exfoliation-based processes are natural bottleneck to fully realize circuit-level and large-scale implementation of 1*T*-$TaS_2$ devices; as for $BaTiS_3$, it represents the class of newly developed phase-change materials and hence, has the lowest level of technological readiness. With only bulk single crystals being used for probing intrinsic physical properties and prototype device demonstration at cryogenic temperatures, $BaTiS_3$-based devices are still far away from achieving optimized individual device performance, let alone any practical higher-level device integration at the moment.

## 2. Coincident electronic and structural phase transitions

Electronic phase transitions often coincide with structural phase transitions in these phase-change materials (Table 1), as well documented in the literature.[69,70] Because these well-defined structural phase transitions are directly tied to the changes of electrical characteristics, elucidating the structural transitions and understanding the associated mechanisms are crucial for developing phase-change-based electronic devices. In the correlated oxide $VO_2$, a monoclinic-to-tetragonal structural transition occurs alongside a pronounced change in resistivity across the metal-to-insulator transition.[41,69] Meanwhile, in low-dimensional CDW chalcogenides such as 1*T*-$TaS_2$ and $BaTiS_3$, the underlying periodic lattice distortions such as star-of-David structural transition (1*T*-$TaS_2$)[53] and in-plane unit cell doubling ($BaTiS_3$)[56] coincide with the abrupt changes in the electrical properties.

Figure 3a shows a representative temperature-dependent conductivity measurement of a $VO_2$ single crystal, which exhibits a resistivity of approximately 10 Ω·cm at room temperature and drops to below $10^{-4}$ Ω·cm at high temperatures, resulting in a change up to five orders of magnitude across the transition.[41] As illustrated in Figure 3b, $VO_2$ adopts a monoclinic crystal structure at room temperature ($a$ = 5.75 Å, $b$ = 4.52 Å, c = 5.38 Å, $\alpha = \gamma = 90^\circ$, $\beta = 122.6^\circ$) with a space group of $P2_1/c$. Above the transition temperature ($T_c$ = 340 K, or 67 °C), the structure becomes tetragonal ($a = b$ = 4.55 Å, $c$ = 2.85 Å, $\alpha = \beta = \gamma = 90^\circ$) with a space group of $P4_2/mnm$, analogous to rutile $TiO_2$.

Despite the clear observations of both structural and metal-to-insulator transitions in $VO_2$, the underlying mechanism has been highly debated. In early days, Goodenough suggested that an antiferroelectric distortion and the formation of V-V zigzag bonds below $T_c$ are responsible for the gap opening and metal-to-insulator transition in $VO_2$.[69] Several experimental observations such as optical phonon softening at the $R$ point of the Brillouin zone also supports this scheme.[70] However, the Peierls' picture does not account for features such as the relatively large bandgap (0.6 eV) and the appearance of additional intermediate phases. Later in 1975, Zylbersztejn and Mott argued that the metal-to-insulator transition in $VO_2$ is triggered when the Hubbard energy ($U$) becomes comparable to the band width.[71] The Mott's criterion for the electronic transition is given as $(n_c)^{1/3}\alpha_H \approx 0.25$, where $n_c$ denotes the critical carrier density and $\alpha_H$ is the Bohr radius. In 2000, Stefanovich *et al*. found that the insulator-to-metal transition in $VO_2$ can be induced by injecting excessing carriers without heating up the lattice to $T_c$, which they construed as strong evidence for the electronic Mott-Hubbard scenario.[72] More importantly, from the perspective of theoretical calculations, phenomena such as broad band gap and the emergence of intermediate phases can only be properly described by introducing an additional Hubbard energy ($U$), through either DFT + $U$ or GW methods.[73-75]

Indeed, there has not been a universal understanding of the $VO_2$ phase transition mechanism, although both Peierls and Mott mechanisms seem to contribute. In 2005, the switching of orbital occupancy across the transition in $VO_2$ was experimentally observed to be directly connected with both the delocalization of electrons and the lattice diction, and hence, an orbital-assisted 'collaborative' Mott-Peierls transition picture was proposed.[76] Moreover, in 2014, Morrison *et al*. reported a photoinduced metal-like monoclinic phase of $VO_2$ from the insulating phase, suggesting that the photoexcitation was only able to rearrange the occupancy of *d* orbitals but insufficient to alter the underlying lattice distortion.[77] More detailed discussions on the mechanism debate of $VO_2$ can be found in several other reviews.[78,79] Note that in Table 1, we label the electron-correlation as the primary mechanism of the metal-to-insulator transition in $VO_2$, as well as in many other correlated oxides.

Charge density waves have been reported in various layered transition metal dichalcogenides (TMDCs) such as 1*T*-$TaS_2$, 2*H*-$TaSe_2$, and 1*T*-$TiSe_2$, in addition to classic quasi-1D metals.[80] The two-dimensional compound $TaS_2$ belongs to the family of TMDCs and crystalizes in different layered structures, including the 1*T* and 2*H* polytypes.[52,81] In 1*T*-$TaS_2$, tantalum (Ta) atoms, octahedrally coordinated by surrounding sulfur atoms, are hexagonally arranged in plane. In 1975, Scruby *et al*. carried out temperature-dependent electron and X-ray diffraction studies[53,82] to reveal three metastable phases in 1*T*-$TaS_2$.

The material exhibits a metallic phase at high temperatures ($a$ = b = 3.36 Å, c = 5.90 Å, $\alpha = \beta = 90^\circ$, $\gamma = 120^\circ$) with the $P\bar{3}m1$ space group, and it switches to an incommensurate CDW (ICCDW) phase below 550 K. The diffraction pattern of the ICCDW phase is dominated by diffuse spots with an incommensurate wave vector $q_{\mathrm{IC}} = 0.283a^*$ .[53] Upon further cooling to approximately 350 K, the wave vector rotates by about 12° toward $q_{\mathrm{NC}} = 0.245a^* + 0.068b^*$,[53] giving rise to a nearly commensurate CDW (NCCDW) phase. Finally, below 180 K, a commensurate CDW (CCDW) phase with a $\sqrt{13}a \times \sqrt{13}b \times 13c$ supercell dominates, which corresponds to a commensurate wave vector of $q_{\mathrm{C}} = (3a^* + b^*)/13 = 0.231a^* + 0.077b^*$.[53]

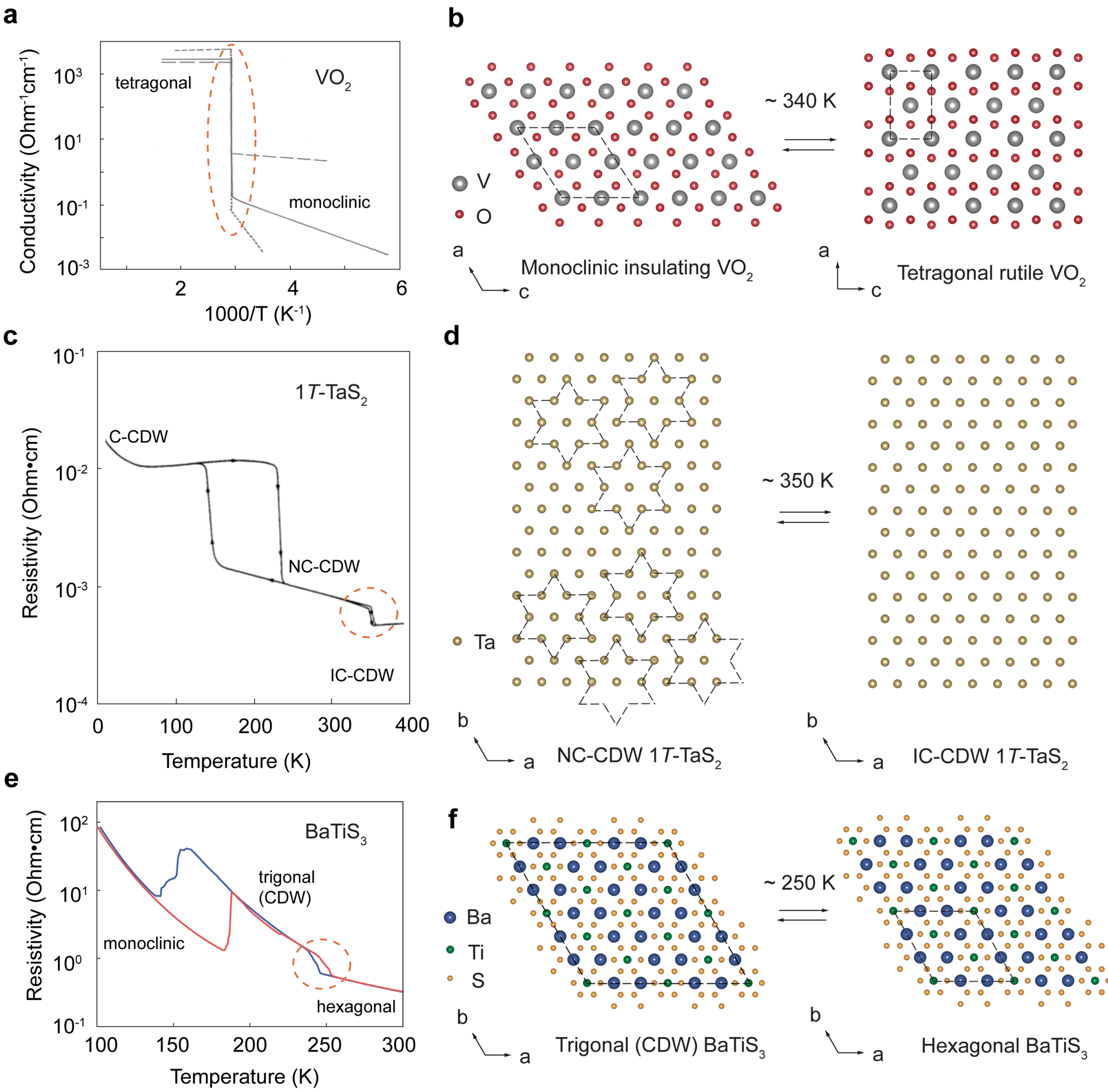

**Figure 3 Coincident electronic and structural phase transitions**. (**a**, **c**, and **e**) Representative temperature-dependent electrical conductivity / resistivity of (**a**) $VO_2$ single crystal, (**c**) 1*T*-$TaS_2$ single crystal, and (**e**) $BaTiS_3$ crystal along the *c*-axis. Figures adapted from Ref. [41] Copyright 1969, American Physical Society, Ref. [52] Copyright 1971, Elsevier, and Ref. [56] Copyright 2023, John Wile & Sons. (**b**, **d**, and **f**) Schematic illustration of the corresponding structural phase transitions in (**b**) $VO_2$, from tetragonal rutile phase to monoclinic phase at ~ 340 K, (**d**) 1*T*-$TaS_2$, from IC-CDW to NC-CDW phase at ~ 350 K, and (**f**) $BaTiS_3$, from hexagonal semiconducting phase to trigonal CDW phase at ~ 250 K.

Accordingly, 1*T*-$TaS_2$ exhibits two pronounced discontinuities in resistivity below 400 K, as illustrated in Figure 3c. Near 350 K, the NCCDW-to-ICCDW transition occurs with minimal thermal hysteresis. In contrast, the low-temperature transition at ~ 200 K between the CCDW and NCCDW phases exhibits a roughly 20-fold increase in resistivity and a large hysteresis spanning tens of kelvins.[52]

In the CCDW phase, the star-of-David clusters form as twelve surrounding Ta atoms displace inward towards a central thirteenth Ta atom within each layer. In 1979, Fazekas and Tosatti proposed that out of the thirteen $5d^1$ electrons, twelve become paired in 'star-bonding' orbitals, leaving the thirteenth electron localized near the cluster center.[83] As a result, only these central electrons contribute to electrical conduction and magnetism. Therefore, a Mott-type localization may occur within the sub-band of these central electrons due to their large separations, which explains the drop in conductivity below 200 K. This electron localization scenario is further supported by Hall measurements performed by Inada *et al*. in 1979,[84,85] which demonstrated that the charge carrier density in the CCDW state is an order of magnitude lower than what would be expected from CDW-induced band gaps alone. As illustrated in Figure 3d, the NCCDW phase also contains star-of-David clusters, albeit arranged in a less uniform pattern. Notably, the structural transition between the NCCDW and ICCDW phases is highly relevant for electrically induced threshold switching and voltage oscillations in $1T$-$TaS_2$ devices operating at room temperature.

Unlike conventional metallic CDW materials, the recently discovered quasi-1D chalcogenide $BaTiS_3$ is a small bandgap $d^0$ semiconductor with an bandgap of approximately 0.3 eV.[86] At room temperature, $BaTiS_3$ adopts a hexagonal crystal structure ($a = b$ =11.7 Å, $c$ = 5.83 Å, $\alpha = \beta = 90^\circ$, $\gamma = 120^\circ$) with a space group of $P6_3cm$.[56] In 2018, Niu *et al*. reported a giant optical anisotropy with a record-high birefringence in single crystals of $BaTiS_3$, sparking significant research interest in this material.[86] Because of its nominally empty conduction band, no phase transition was initially expected for $BaTiS_3$, even though its $d^1$ counterpart, $BaVS_3$, is considered an archetypical CDW system with a magnetic transition.[87,88] In 2023, Chen *et al*. experimentally demonstrated the presence of a CDW phase and a series of transitions in $BaTiS_3$ using electrical transport measurements and temperature-dependent synchrotron X-ray diffraction.[56] Upon cooling from room temperature to approximately 240 K, a structural transition takes place with titanium atoms displacing in *a*-*b* plane, which leads to a lattice unit cell doubling ($a = b$ =23.3 Å, $c$ = 5.84 Å, $\alpha = \beta = 90^\circ$, $\gamma = 120^\circ$) and hence a new CDW phase with a space group of $P3c1$, as illustrated in Figure 3f. Key evidence for the CDW includes the emergence of weak superlattice reflections in the diffraction pattern and the corresponding resistivity anomalies observed from transport measurements.[56] Further cooling to 130 K causes these superlattice peaks to disappear while a new set of reflections associated with a smaller $\frac{2}{\sqrt{3}} \times \frac{2}{\sqrt{3}}$ unit cell ($a = b$ =13.4 Å, $c$ = 5.82 Å, $\alpha = \beta = 90^\circ$, $\gamma = 120^\circ$) emerges. This observation indicates a suppression of the CDW phase *via* the structural transition from $P3c1$ to $P2_1$. Consistently, transport measurements reveal two hysteretic transitions in the 150 K – 190 K and 245 – 255 K ranges, respectively, as shown in Figure 3e.[56]

The underlying mechanisms driving these phase transitions in a gapped semiconductor such as $BaTiS_3$ can be complicated. Because its Fermi level lies within the bandgap, these is no conventional concept of Fermi surface; thus, the nesting mechanism associated with quasi-1D CDW metals does not apply.[56] Hall measurements reveal a low carrier concentration of $\sim 1.1 \times 10^{18}$ $cm^{-3}$ at room temperature, which further drops to less than $10^{15}$ $cm^{-3}$ at 100 K.[56] In such a non-degenerate system with dilute concentration of electrons, the role of electron-electron interaction in $BaTiS_3$ can be nontrivial, unlike most metallic or semi-metallic CDW compounds. Therefore, Chen *et.al*. suggested that both electron-lattice coupling and non-negligible electron-electron interactions could contribute to the observed CDW order and phase transitions in semiconducting $BaTiS_3$.[56]

## 3. Materials synthesis

This section provides an overview of the synthesis routes employed for the three phase-change material systems covered in this review ($VO_2$, 1*T*-$TaS_2$, and $BaTiS_3$). Owing to their distinct chemistries, physical properties, and different levels of research focus since their initial discoveries, the methods used to synthesize these materials vary significantly. In general, single-crystal forms of materials are being used in early days of research for studying their intrinsic physical properties and the demonstration of prototype devices. By contrast, large-area, high-quality thin film growth is usually crucial for realizing any practical electronic device applications. Table 2 summarizes details of representative material synthesis methods for each material and the associated transport properties.

In 1959, Morin conducted the first electrical transport study of various oxides showing MIT, such as $Ti_2O_3$, VO, $V_2O_3$, and $VO_2$.[35] In his work, single crystals of vanadium oxides were synthesized *via* a hydrothermal process. The details of the crystal growth using this method can be found in a subsequent book chapter by Laudise and Nielsen.[89] At that time, however, the single-crystal samples were on the order of 0.1 mm in size, which were too small for standard four-point measurements. Consequently, the observed electrical conductivity change across the MIT in $VO_2$ was limited to merely two orders of magnitude,[35] likely due to significant contact resistances in two-probe geometry. Thereafter, a variety of advanced synthesis methods have been developed to produce high-quality and large-sized $VO_2$ single crystals. For instance, in 1969, Ladd and Paul employed a molten-flux technique to grow millimeter-scale $VO_2$ single crystals using $V_2O_5$ as flux.[90] These crystals showed a $10^5$-fold change in resistivity near 340 K, which is widely considered the benchmark performance for $VO_2$.[41,90] In 1971, Nagasawa adopted a chemical vapor transport (CVT) method[91] to obtain single crystals of vanadium oxides using $TeCl_4$ as a transport agent,[92] and in 1972, Reyes *et al*. succeeded in synthesizing doped $VO_2$ single crystals *via* an iso-thermal flux evaporation method.[93]

Research efforts on synthesizing $VO_2$ thin films have been predominantly centered on using reactive magnetron sputtering[94,95] and pulsed laser deposition (PLD),[96-99] although other methods including molecular beam epitaxy (MBE),[100,101] metal-organic chemical vapor deposition (MOCVD),[102] and sol-gel routes[103,104] have also been investigated. In 1967, Fuls *et al*. synthesized $VO_2$ thin films by reactive sputtering from a vanadium target in an argon atmosphere with a controlled oxygen partial pressure.[94] These films, grown at 400˚C on sapphire substates, exhibited a highly orientated monoclinic phase at room temperature.[94] From late 1980s, PLD emerged as a versatile technique for synthesizing high-quality oxide thin films (Figure 4a). Its compatibility with relatively high oxygen pressures makes it particularly suitable for depositing stoichiometric oxides. In 1994, Kim and Kwok demonstrated high-quality $VO_2$ thin films on (0001) and $(10\bar{1}0)$ sapphire substrates by PLD, using pressed $V_2O_3$ powder as the target.[96] More recently, Zhang *et al*. reported wafer-scale $VO_2$ growth on sapphire substrates (Figure 4b, left) with a large resistance jump of ~ $10^4$ through a hybrid-MBE approach. The right panel of Figure 4b shows a representative XRD scan of such a $VO_2$ film.[105]

It is important to note that the transport properties of $VO_2$ such as the transition temperature, magnitude of resistivity change, and the width of the hysteresis loop depend strongly on the material synthesis method and growth conditions, as detailed in Table 2. For instance, a typical $VO_2$ thin film grown by RF sputtering exhibits a three-order-of-magnitude resistivity changes and a 10 K hysteresis.[106] In contrast, high-quality $VO_2$ thin films synthesized *via* pulsed laser deposition can attain a resistivity change of up to five orders of magnitude with a 1 K hysteresis,[96] comparable to that observed in bulk single crystals,[90,107] yet its transition temperature is reduced to ~ 330 K. Moreover, it has been reported that in the thin film limit of $VO_2$ (10-15 nm thick), the epitaxial strain, which is induced by the differences of the lattice constants between the film and the $TiO_2$ (001) substrate, could dramatically reduce the transition temperature to ~ 300 K.[110]

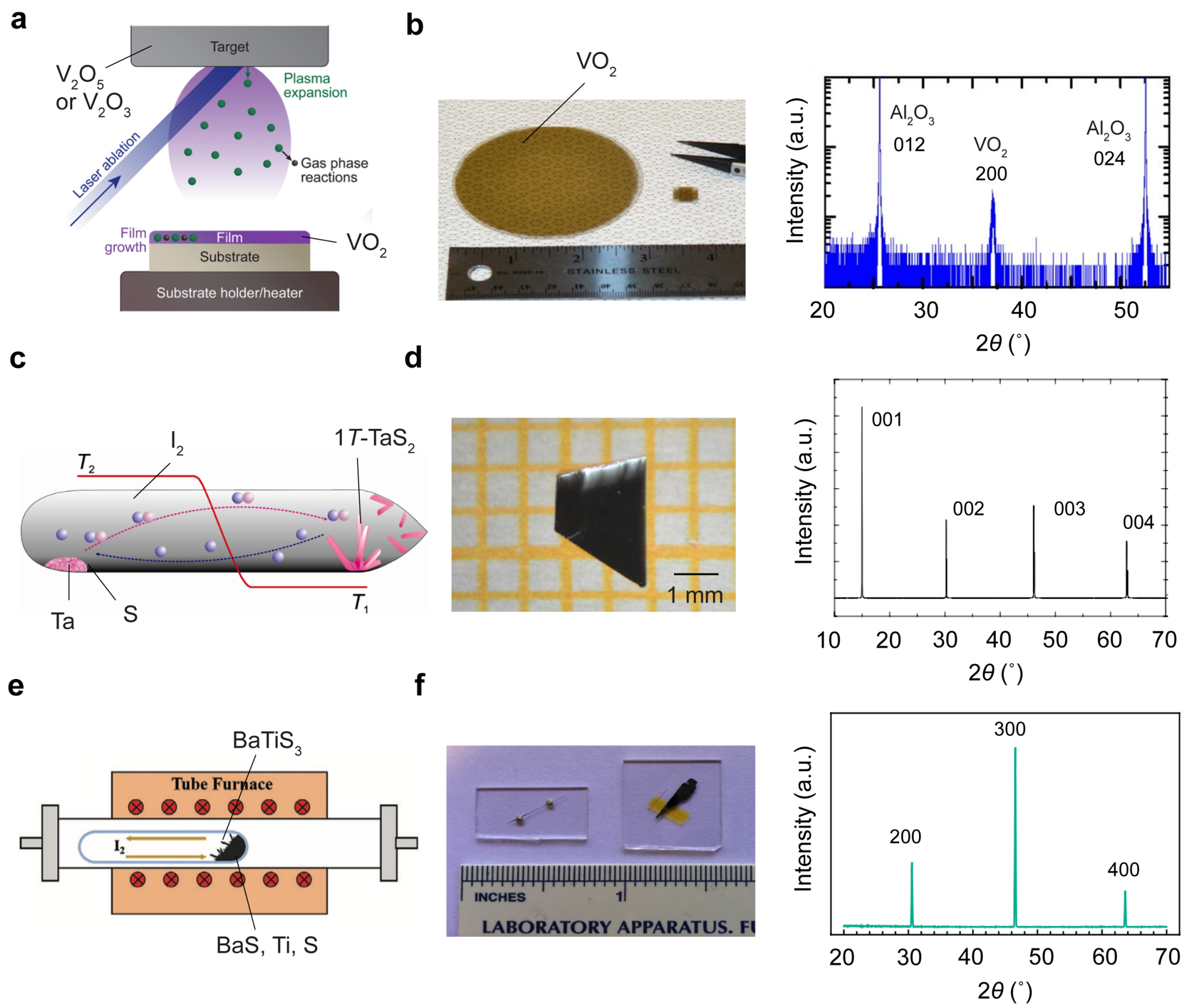

**Figure 4 Material synthesis.** (**a**) Schematic illustration of a PLD system for epitaxial thin-film growth of $VO_2$ using a $V_2O_5$ or $V_2O_3$ target. Figure adapted from Ref. [99] Copyright 2023, Royal Society of Chemistry. (**b**) Optical image (left) and a representative XRD scan (right) of $VO_2$ thin films grown on sapphire substrates by hybrid-MBE. Figures adapted from Ref. [105] Copyright 2015, Springer Nature. (**d**) Schematic illustration of 1*T*-$TaS_2$ single crystal growth using a CVT method. The schematic is adapted from Ref. [91] Copyright 2013, Intech Open. (**e**) Optical image (left) and the corresponding out-of-plane XRD scan (right) of a large-sized 1*T*-$TaS_2$ single crystal. Figures adapted from Ref. [108] Copyright 2015, Springer Nature. (**g**) Schematic illustration of $BaTiS_3$ single crystal synthesis using a vapor transport method. Figure adapted from Ref. [109] Copyright 2022, John Wiley & Sons. (**f**) Optical image (left) of representative as-grown $BaTiS_3$ crystals with needle-like and plate-like morphologies and out-of-plane XRD scan (right) of a $BaTiS_3$ plate with *a*- and *c*-axes in plane. Figures adapted from Ref. [86] Copyright 2018, Springer Nature.

Single crystals of 1*T*-$TaS_2$ are commonly obtained by the CVT technique, like many other layered TMDCs, as illustrated in Figure 4c.[91] A comprehensive review of this method can be found in the book by Schäfer (1964).[111] By 1969, Wilson and many other researchers had succeeded in synthesizing TMDCs single crystals (including $TaS_2$) with dimensions up to a centimeter.[81] Iodine ($I_2$) serves as the most widely used transport agent, although bromine and chlorine are also occasionally employed. Typically, ~ 1 mg/cm$^3$ of $I_2$ is placed in a sealed quartz ampule to balance an efficient reaction rate with minimal unintentional iodine incorporation. Figure 4d (left) displays an optical image of a representative bulk 1*T*-$TaS_2$ crystal (2 mm × 3 mm), with the right panel illustrating an XRD scan of the same sample.[108] In 1971, Thompson *et al*. obtained 1*T*-$TaS_2$ crystals by quenching sealed ampules from ~ 950˚C, following a CVT crystal growth process using pre-reacted 2*H*-$TaS_2$ powder as the starting materials and $I_2$ as the transport agent.[52] Notably, the 1*T* polymorph of $TaS_2$ is thermodynamically stable at temperatures above 777˚C (~1050 K) in its phase diagram, and therefore, growths with slow cooling will lead to 2*H*-$TaS_2$, which is metallic and becomes superconducting below 0.8 K.[83] With sufficiently large crystal sizes, Thompson *et al*. further performed transport measurements up to 400 K using a van der Pauw geometry, revealing the intrinsic electrical transport behavior of bulk 1*T*-$TaS_2$ featuring two metal-to-insulator phase transitions[52]. Those two transitions were later assigned as the CCDW-to-NCCDW and NCCDW-to-ICCDW transitions, respectively. Due to the upper limit of the measurement temperature, the higher-temperature metallic phase was not captured in that study.[52]

**Table 2 Material synthesis and transport properties of $VO_2$, 1*T*-$TaS_2$, and $BaTiS_3$.**

| Material | Material Synthesis | | | Transport Properties | | | Year | Ref. |
|---|---|---|---|---|---|---|---|---|
| | Synthesis Method | Substrate | Growth Temperature | Transition Temperature | Resistivity Change | Hysteresis Window | | |
| $VO_2$ | Flux (single crystal) | | 1000˚C | 340 K | $10^5$ | 0.5-1 K | 1969 | 41,90 |
| | Sputtering | Sapphire | 400˚C | 340 K | $10^3$ | 10 K | 1967 | 94 |
| | PLD | Sapphire $(10\bar{1}0)$ | 630˚C | 328 K | $10^5$ | < 1 K | 1994 | 96 |
| | PLD | $TiO_2$ (001) | 370 ˚C | 300 K | $10^3$ | 10 K | 2002 | 110 |
| | Sol-gel | Sapphire $(10\bar{1}0)$ | 400-500˚C | 340 K | $10^4$ | 10 K | 2005 | 103 |
| 1*T*-$TaS_2$ | CVT (single crystal) | | 950˚C | 350 K | ~2 | 5 K | 1971 | 52 |
| | CVD | *h*BN | 850˚C | 345 K | ~2 | 30 K | 2018 | 112 |
| $BaTiS_3$ | CVT (single crystal) | | 1050˚C | 250 K | 2 | 10 K | 2023 | 56 |
| | Flux (single crystal) | | 1050˚C | 230 K | 2-3 | 20 K | 2024 | 113 |
| | PLD | $SrTiO_3$ | 700˚C | N/A | N/A | N/A | 2024 | 114 |

Thus far, most existing thin 1*T*-$TaS_2$ devices have been fabricated by mechanically exfoliating bulk crystals. While this method yields high-quality flakes, it restricts both sample size and fabrication throughput. As a result, establishing large-area, controllable, and high-quality thin-film synthesis of 1*T*-$TaS_2$ is crucial for advancing both fundamental studies – especially at the thin limit – and practical device applications. In 2016, Fu *et al*. demonstrated a chemical vapor deposition (CVD) approach to grow 1*T*-$TaS_2$ thin flakes of varying thickness on $SiO_2$/Si substrates, using $TaCl_5$ and sulfur powder as precursors under an $H_2$/Ar atmosphere at 1093 K.[115] Later in 2018, Wang *et al*. reported the electrical measurements from CVD-grown 1*T*-$TaS_2$ thin films on *h*BN, where the NCCDW-to-ICCDW transition features a pronounced thermal hysteresis of 30 K and the low-temperature CCDW-to-NCCDW transition is largely suppressed, deviating from the transport properties of 1*T*-$TaS_2$ crystals.[112] Alternatively, Lin *et al*. developed an MBE growth method for $TaS_2$ on graphene-terminated 6*H*-SiC (0001) substrates.[116] At a growth temperature of ~700˚C, both 1*T* and 2*H* phases of $TaS_2$ were obtained, and the authors suggested that higher substate temperatures favor the formation of the 1*T* phase, however, no transport properties were reported.[116]

In stark contrast to $VO_2$ and 1*T*-$TaS_2$, $BaTiS_3$ has received far less attention as a phase-change material, and its single-crystal form was not available prior to 2018, despite the synthesis and structural characterization of $BaTiS_3$ powders dating back to 1957.[117] In 1996, Imai *et al.* measured the specific heat of pressed $BaTiS_3$ powder from 1.4 K to 300 K, but they did not observe any anomaly indictive of phase transitions.[118] The absence of observable transitions remains puzzling; one plausible hypothesis is that powders of $BaTiS_3$ contain significantly more point and extended defects than single crystals, thereby suppressing the transitions. Further thermodynamic studies of large, high-quality $BaTiS_3$ single crystals are expected to clarify this issue.

In 2018, Niu *et al.* reported the first successful growth of $BaTiS_3$ single crystals using a vapor-phase growth approach with $I_2$ as the transport agent.[86] Unlike many conventional CVT processes where source materials are transported towards the other end of the ampule, clusters of needle-like $BaTiS_3$ crystals (typically < 50 μm in both width and thickness) were observed to directly grow out of $BaTiS_3$ powder,[86] as subsequently confirmed by Yang *et al.* in 2022[109] (Figure 4e). Beyond these needle-like morphologies, Niu *et al.* also obtained thin platelets of $BaTiS_3$ with *a*- and *c*- axes in plane that are suitable for optical studies, as shown in Figure 4f (left). The right panel of Figure 4f illustrates a representative out-of-plane X-ray diffraction scan of such a plate-like $BaTiS_3$ crystal. These samples facilitated the full characterization of $BaTiS_3$'s giant optical anisotropies, with a record-high birefringence of up to 0.76 in the mid- to long-infrared range.[86] Subsequently, Zhao *et al.* synthesized thin flakes of 001-type $BaTiS_3$ flakes with *a*- and *b*-axes in plane through a similar CVT route, albeit with slightly modified conditions.[119] Recently, Chen *et al.* developed a molten-flux approach using either potassium iodide (KI) or a mixture of barium chloride ($BaCl_2$) and barium iodide ($BaI_2$) to grow $BaTiS_3$ crystals.[113] The KI-based flux approach yielded crystals of dimensions up to a centimeter in length and 500 μm in both width and thickness, whereas the $BaCl_2$-$BaI_2$ flux method produced plate-like, (001)-oriented $BaTiS_3$ crystals up to 200 μm thick. These flux-grown crystals exhibit substantially larger volumes than those obtained *via* vapor transport, while preserving the material's intrinsic optical and electronic properties.[113]

The development of high-quality thin-film growth of $BaTiS_3$ still faces substantial challenges. Despite the successful demonstration of $BaTiS_3$ thin films using magnetron sputtering[120] and solution processing[121] for optoelectronic applications, efforts towards realizing epitaxial thin films growth of $BaTiS_3$ for leveraging its intrinsic optical anisotropy and phase-change properties have, to date, been attempted exclusively *via* PLD approaches developed for complex chalcogendies.[114,122,123] In 2022, Surendran *et al*. demonstrated the quasi-epitaxial synthesis of $BaTiS_3$ on single-crystalline $SrTiO_3$ substrates at approximately 700˚C, using an Ar/$H_2S$ (5%) background atomosphere.[114] X-ray diffraction measurements revealed a pronounced out-of-plane texture in these films, although no clear in-plane epitaxial relationship between thin film and the substrate was observed.[114] More recently, Surendran *et al*. developed a hybrid PLD strategy that employs an organosulfur precursor as the background sulfur source, substituting the chemically aggressive Ar/$H_2S$ environment.[122] $BaTiS_3$ films grown on $SrTiO_3$ using this hybrid PLD method exhibited substantially improved surface and interface smoothness, while retaining crystallinity comparable to that achieved by conventional PLD using Ar/$H_2S$.[122] It is worth noting that proper defect control in growth condition optimization, particularly of sulfur vacancies, can be critical in realizing high-quality epitaxial $BaTiS_3$ thin films, as has been demonstrated on another ternary chalcogenide $BaZrS_3$.[124,125] The scarcity of lattice-matched and chemically compatible substrates can be another limiting factor. At the moment, regular perovskite oxide substates such as $SrTO_3$, $LaAlO_3$, and $SrLaAlO_4$ are still the only few available options for complex chalcogenide synthesis.[114,122,123] Therefore, achieving wafer-scale complex chalcogenide single crystals through advanced synthesis techniques such as top-seeded solution growth or Bridgeman method, represents a vital step towards realizing high-quality chalcogenide thin films and their heterostructures in the future.

## 4. Device fabrication

This section provides an overview of the recent advancements in electrical contact and device fabrication processes for three phase-change materials – $VO_2$ (thin films), 1*T*-$TaS_2$ (mechanically cleaved thin flakes), and $BaTiS_3$ (micro-scale bulk crystals). Owing to the different susceptibilities of these materials' phase transitions to external factors such as oxidation and strain, researchers have developed and implemented various strategies to preserve intrinsic transport properties and achieve the desired device performance.

Fabrication processes for $VO_2$ tend to be the most straightforward among the three materials, largely owing to its excellent air / thermal stability and the well-developed processes for synthesizing high-quality thin films whose monoclinic-to-tetragonal phase transition is well-preserved.[94,96] Hence, conventional cleanroom micro-fabrication techniques, such as photolithography, etching and metal deposition, can be readily applied. Device geometries of $VO_2$ such as multi-terminal Hall bars or linear bars are often patterned through dry etching approaches such as reactive ion etching (RIE) or ion milling for proper transport measurements.[126-128] Simpler two-terminal geometries are generally sufficient for neuronal oscillator devices (Figure 5a and 5b), where a short channel length proves crucial to achieving low switching voltages and high oscillation frequencies.[126,129,130] When device dimensions must be reduced beyond the limits of standard lithography techniques, an out-of-plane $VO_2$ device geometry, where the channel dimension is equivalent to the film thickness (on the order of tens of nanometers), is often employed.[131,132] In such configurations, $VO_2$ films are typically deposited on high-temperature-compatible conductive substrates, including Pt, TiN, and $SrRuO_3$, before putting down top electrodes.[97,131,132] In 2015, Mian *et al*. reported a record-high oscillation frequency of 9 MHz from a $VO_2$/TiN vertical device,[131] as illustrated in Figure 5c. Additional optimization of metal-to-$VO_2$ contacts is expected to further improve the device performance.

1*T*-$TaS_2$ exhibits a layered structure bonded *via* weak van der Waals interactions, and hence, it can be mechanically exfoliated down to a few tens of nanometers or thinner for device fabrication. Notably, there has been a long history of research on exfoliation and transfer of layered TMDCs,[81,133] dating back well before the discovery of single-layer graphene in 2004.[134,135] For instance, in 1962, Frindt and Yoffe produced ultra-thin molybdenum disulfide ($MoS_2$) flakes (< 10 nm) by mechanical cleavage for optical and electron diffraction studies.[133] By 1969, Wilson and Yoffe were able to prepare thin specimens (<100 nm) of various TMDCs by repeated cleaving on adhesive tapes.[81] More recently, atomically thin 1*T*-$TaS_2$ has re-attracted enormous research interest, revealing novel phenomena such as gate-tuned phase transition,[108] electrical oscillations,[62,63] and memristive switching.[54,136] These advances have been facilitated by the fast-evolving developments of exfoliation and transfer techniques for 2D materials, wherein thin flakes of 1*T*-$TaS_2$ with thickness down to monolayer can be isolated and integrated into devices. E-beam lithography is often adopted to pattern fine features at the micrometer scale or smaller, and to accommodate the specific alignment requirement for each flake. However, due to high processing temperatures of most ebeam resists (e.g., ~ 180˚C baking temperature for PMMA), additional care needs to be taken to avoid material degradation when fabricating 1*T*-$TaS_2$ based devices, particularly at their thin limits.

In 2015, Yu *et al*. observed that below 10 nm, the CCDW-to-NCCDW transition in 1*T*-$TaS_2$ is suppressed, whereas the higher-temperature NCCDW-to-CCDW transition at ~ 350 K persists till ~ 4 nm, below which it also disappears.[108] Debate arises regarding whether the apparent absence of CDW phase transitions in ultrathin, unprotected 1*T*-$TaS_2$ flakes arises from intrinsic dimensional effects or extrinsic factors such as oxidation, considering the thin material's propensity to degradation in air (Figure 5d). Soon after Yu's work, Tsen *et al*. demonstrated that ultrathin 1*T*-$TaS_2$ flakes, protected by hexagonal boron nitride (*h*BN), retained the NCCDW-to-CCDW transition at thickness down to 4 nm through both electron diffraction and transport measurements (Figure 5e).[137] They even showed evidence of the NCCDW phase in the 2 nm sample using electron diffraction, although the transition to the CCDW was not observed at this thin limit.[137] These findings suggest that oxidization, rather than purely dimensional effects, may be responsible for the absence of charge order in ultrathin, unprotected 1*T*-$TaS_2$ flakes.[137] Therefore, *h*BN encapsulation strategy is usually employed to preserve the intrinsic device performance of 1*T*-$TaS_2$ when working with ultrathin samples[108,137]. On the other hand, although mechanical exfoliation and *h*BN encapsulation strategies are suitable for laboratory-scale research, it is not practical for large-area device fabrication. Consequently, further efforts need to be made on high-quality thin-film synthesis and device encapsulation of 1*T*-$TaS_2$, where intrinsic electrical switching properties can be well-preserved.[115,116]

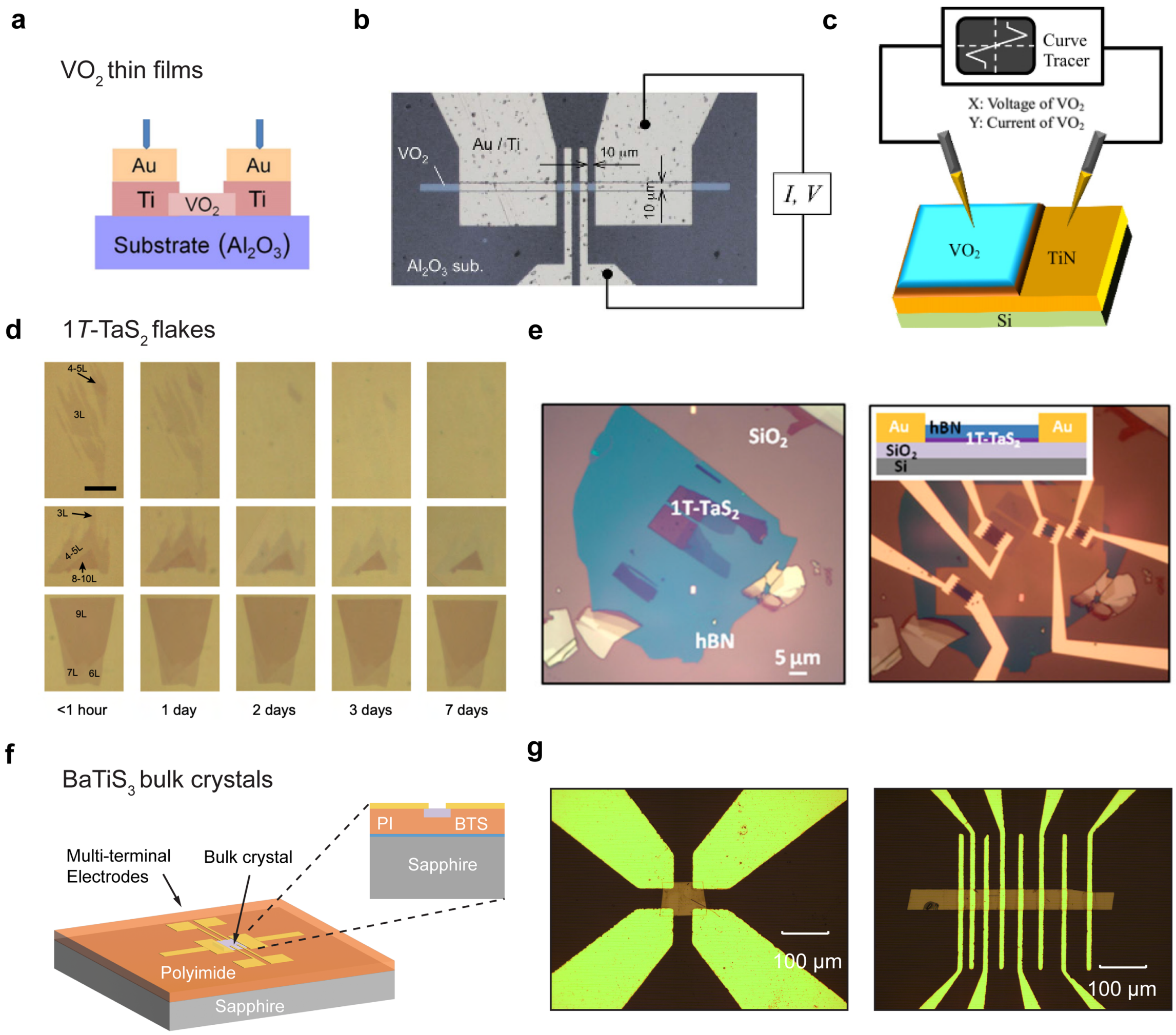

**Figure 5 Device Fabrication.** (**a**) Cross-sectional schematic of a two-terminal planar $VO_2$ thin film device. Figure adapted from Ref. [130] Copyright 2011, Elsevier. (**b**) Optical microscopic image of a four-probe $VO_2$ film device fabricated through dry etching. Figure adapted from Ref. [126] Copyright 2008, AIP Publishing. (**c**) Schematic illustration of an out-of-plane $VO_2$ film device using TiN as the bottom electrode. Figure adapted from Ref. [131] Copyright 2015, AIP Publishing. (**d**) Optical microscopic images of few-layer 1*T*-$TaS_2$ flakes showing the degradation progress in air. Figures adapted from Ref. [108] Copyright 2015, Springer Nature. (**e**) Optical microscopic images of device fabrication processes for mechanically exfoliated 1*T*-$TaS_2$ flakes on a $SiO_2$/Si substrate, including *h*BN encapsulation (left) and metal deposition (right) steps. The inset to the right shows a cross-sectional schematic of the device. Figures adapted from Ref. [137] Copyright 2015, National Academy of Science. (**f**) Schematic illustration of a multi-terminal $BaTiS_3$ bulk device fabricated through polymeric planarization and regular photolithography processes. Figure adapted from Ref. [138] Copyright 2022, ACS Publications. (**g**) Optical microscopic images of $BaTiS_3$ bulk devices with a van der Waals (left) and a transmission line method (TLM, right) contact geometry. Figures adapted from Ref. [56] Copyright 2023, John Wiley & Sons.

For $BaTiS_3$, like many other small, non-exfoliable crystals, establishing high-quality multiterminal electrodes on them has proven challenging due to their limited size. Although Niu *et al*. has demonstrated the synthesis of large, thin $BaTiS_3$ platelets with lateral dimensions of several millimeters *via* a vapor transport method in his original optical anisotropy study in 2018,[86] the yield of such large samples suitable for manual bonding remains low, due to the large number of nucleation sites during growth. Indeed, most $BaTiS_3$ crystals grown by CVT methods exhibit a needle-like shape that is typically tens of microns in thickness and width, and up to several millimeters along the chain axis, as confirmed by several recent reports.[86,109] In 2022, Chen *et al*. addressed this challenge by adapting a polymeric planarization technique, initially developed for handling micro-scale GaAs-based devices,[139,140] to fabricate multiterminal electrical contacts on small, bulk $BaTiS_3$ crystals for electrical transport studies.[138,141] As depicted in Figure 5f, the crystals are first embedded in a low-stress polymeric matrix (e.g., polyimide) to achieve a planar top surface, enabling direct application of standard lithography and microfabrication processes for forming electrodes with desired geometries.[138] The choice of a low-stress polymer minimizes extrinsic thermal strain effects at low temperatures, which preserves intrinsic transport characteristics of $BaTiS_3$ that are highly susceptible to strains.[56,138] These prototype $BaTiS_3$ single-crystal devices (Figure 5g) exhibited clear signatures of phase transitions and enabled proof-of-concept demonstrations of electrical resistive switching and voltage oscillations.[34,56] Further research and development on high-quality $BaTiS_3$ thin-film synthesis that features intrinsically robust or tunable phase transitions, will help streamline its device fabrication and pave the way toward practical electronic devices applications of this emerging phase-change material.

## 5. Transport properties

Temperature-dependent electrical transport measurements are among the most important characterizations for phase-change materials. Upon varying the temperature, anomalies in electrical resistivity such as metal-to-insulator transitions can emerge near specific transition temperatures, typically reflecting an underlying structural transition and corresponding modification in the electronic structure. In addition to the classic metal-to-insulator transitions observed in the correlated oxide $VO_2$,[35] both the CDW systems 1*T*-$TaS_2$ and $BaTiS_3$ display pronounced resistivity changes across the transitions.[52,56] It is important to note, however, that the metal-to-insulator transition is not a universal feature in all real CDW materials, despite the prediction of Peierls' theory for ideal one-dimensional metals.[142,143] Detailed temperature-dependent Hall measurements (Figure 6a), which probe the evolution of carrier concentration and mobility, further provide important insights into the mechanisms governing these phase transitions.

As detailed in Figure 3a, single crystals of $VO_2$ undergo dramatic change in resistivity of up to five orders of magnitude across the transition.[41] In 1973, Rosevear and Paul investigated magneto-transport properties of both semiconducting and metallic phases in $VO_2$ single crystals using an oscillating AC magnetic field,[144] as shown in Figure 6b. At the insulator-to-metal transition, the carrier concentration extracted from the Hall measurements increased by roughly a factor of $5 \times 10^4$, whereas the Hall mobility changed from about 0.5 $cm^2/(V \cdot s)$ to 0.35 $cm^2/V \cdot s$.[144]

1*T*-$TaS_2$ exhibits two pronounced hysteretic transitions in resistivity below 400 K, as illustrated in Figure 3c. The exact hysteresis width varies among samples, which is presumably associated with impurities, structural defects or other forms of disorder. In fact, the earliest electrical transport study on 1*T*-$TaS_2$ by Thompson *et al*. in 1971 reported hysteresis width as large as 100 K for the CCDW-to-NCCDW transition.[52] In 1980, Inada *et al*. found an even larger resistivity hysteresis of 150 K in 1*T*-$TaS_{1.2}Se_{0.8}$ single crystals, which was attributed to the increased disorder from selenium alloying.[85] Moreover, a substantial rise in resistivity is often observed at temperatures below ~ 50 K. In 1977, Di Salvo and Graebner attributed this low-temperature increase in resistivity to an extrinsic Anderson localization effect by random impurities or defect potentials.[145] Hall measurements of 1*T*-$TaS_2$ provide further insights into its transport processes by revealing both the carrier type and concentrations, as demonstrated by Inada *et al*. in 1979 (Figure 6c).[84] The high-temperature metallic phase transforms into the ICCDW phase, where the Fermi surface is broken into fragments as a result of CDW formation. In this ICCDW phase, the residual Fermi surface fragments still give rise to a high carrier concentration of ~ $10^{22}$ $cm^{-3}$ and the majority carriers are electrons.[84] Upon crossing the NCCDW transition near 350 K, the carrier concentration drops by about 30%, and continues to decrease to ~ $3 \times 10^{21}$ cm$^{-3}$ down to ~ 200 K. At the NCCDW-to-CCDW transition upon cooling, a discontinuous change in resistivity accompanies a sign reversal of the Hall coefficient ($R_H$).[84] The system enters a semimetallic or semiconducting region with its majority carrier type switched to p-type at carrier concentrations of ~ $10^{19}$ cm$^{-3}$. A characteristic peak of $R_H$ around 30-50 K is further postulated that, below that temperature range, the material becomes a semiconductor of the impurity type.[84]

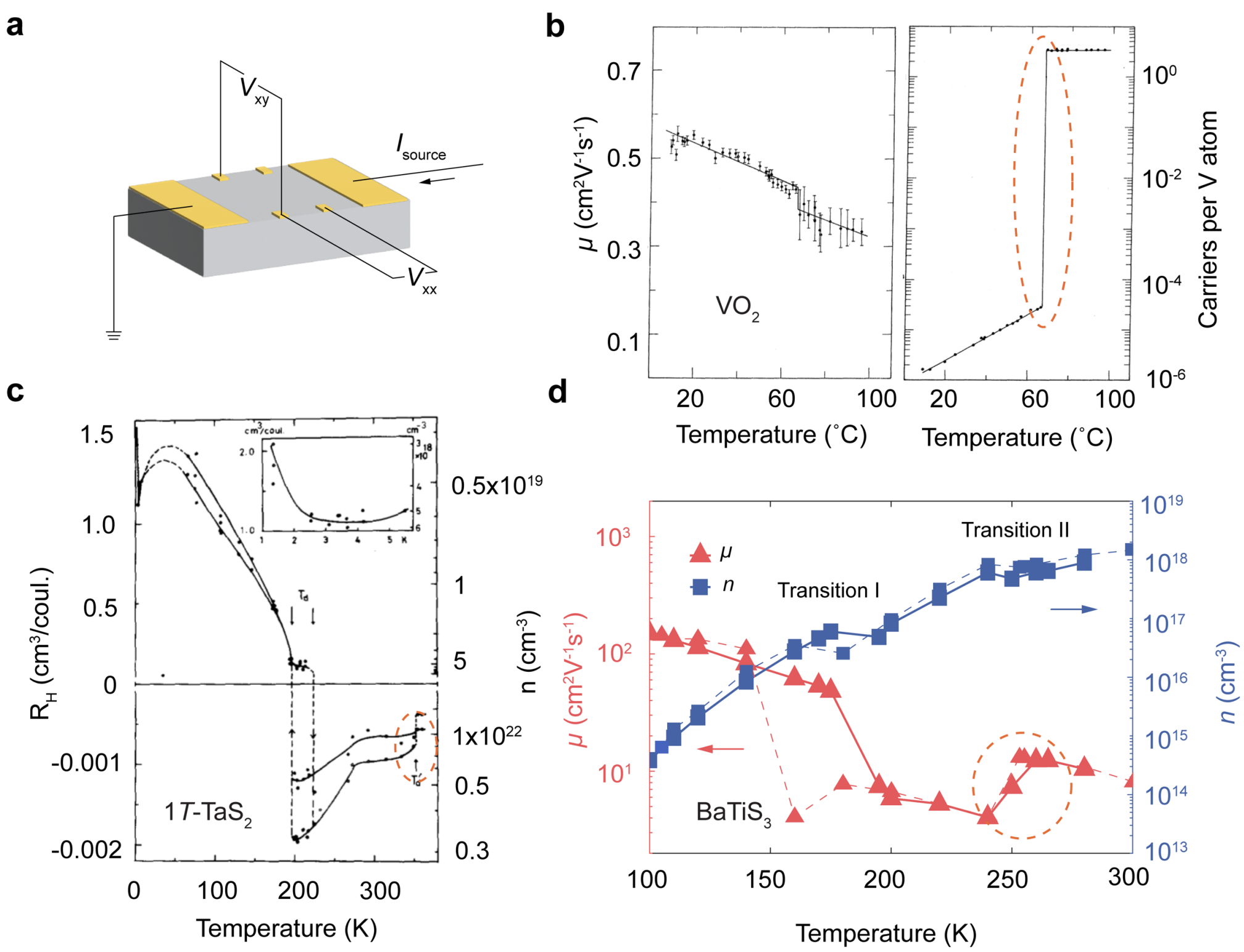

**Figure 6 Transport properties.** (**a**) Schematic illustration of Hall measurements. (**b**) Temperature evolution of Hall mobility (left) and carrier concentration (right) across the MIT phase transition in $VO_2$. Figures adapted from Ref. [144] Copyright 1973, American Physical Society. (**c**) Temperature dependence of Hall coefficient and carrier concentration across the transitions. Figure adapted from Ref. [84] Copyright 1979, Elsevier. (**d**) Temperature evolution of mobility and carrier concentration of a $BaTiS_3$ crystal along the *c*-axis. The associated temperature dependent electrical resistivity measurement is shown in Figure 3e. Figure adapted from Ref. [56] Copyright 2023, John Wiley & Sons.

Moreover, by simply reducing the thickness of 1$T$-$TaS_2$, researchers have significantly modulated its transport characteristics compared with bulk samples. In 2015, Yu *et al.* systematically investigated 1$T$-$TaS_2$ thin flakes with varying thicknesses down to ~ 2 nm, observing a pronounced broadening in the hysteresis width of the low-temperature NCCDW-to-CCDW transition as the thickness decreased from ~ 100 nm to ~ 10 nm, alongside a less abrupt overall resistivity change.[108] Below 10 nm, the CCDW-to-NCCDW transition is suppressed, whereas the higher-temperature NCCDW-to-CCDW transition at ~ 350 K persists till ~ 4 nm, below which it also disappears.[108] Notably, Tsen *et al*. demonstrated that, by simply encapsulating ultrathin 1*T*-$TaS_2$ flakes with hexagonal boron nitride (*h*BN), NCCDW-to-CCDW transition can be retained in samples with thicknesses down to 4 nm, suggesting that oxidization plays an important role in suppressing the charge order in ultrathin, unprotected 1*T*-$TaS_2$ flakes.[137]

In contrast, $BaTiS_3$ was not known to exhibit any phase transition until recently. In 2023, Chen *et al*. identified two distinct transitions in single-crystal $BaTiS_3$ from its non-monotonic, hysteretic transport behavior,[56] as shown in Figure 3e. Upon cooling, the system undergoes a transition from the ambient semiconducting phase to the CDW phase at ~ 240 K ("Transition II") with a clear resistivity jump. On further cooling, the resistivity continues to rise till ~ 150 K, whereupon another transition ("Transition I") drives the system into a high-conductivity state (or high-mobility state), marked by a sharp resistivity drop. The hysteresis spans over 40 K for Transition I and about 10 K for Transition II.[56] Structural characterizations revealed weak superlattice reflections indictive of a periodic lattice distortion emerging below Transition II, and these are subsequently suppressed at Transition I.[56] Below 100 K, the resistivity increases rapidly and becomes too large to measure reliably using standard AC transport techniques.

Hall measurements of $BaTiS_3$ single crystals further clarify its transport mechanisms.[56] With electron as majority carrier, $BaTiS_3$ has a room-temperature carrier concentration of ~ $1.1 \times 10^{18}$ $cm^{-3}$, which is among the lowest reported for known CDW compounds, confirming its nondegenerate nature.[56] As illustrated in Figure 6d, the carrier concentration decreases monotonically upon cooling, reaching ~ $10^{15}$ $cm^{-3}$ by 100 K, without any abrupt changes in magnitude or sign across either transition.[56] Instead, Hall mobility undergoes a significant drop and then a substantial rise at Transition II and I, respectively, suggesting that the modulation of $BaTiS_3$'s resistance stems largely from variations in mobility.[56] This contrasts notably with $VO_2$ and 1*T*-$TaS_2$, where large carrier concentration changes are responsible for the resistance jumps across phase transitions.[84,144]

## 6. Electrically driven volatile threshold switching and mechanisms

In addition to varying system temperature, adjacent phases in phase-change materials can be switched electrically by applying either voltages or currents.[32-34,146] Such electrically controlled resistive switching holds unique advantages over other modulating parameters such as temperatures, high pressure, and optical pulses, owning to the relative ease of implementation in electronic devices. Depending on the inherent nature of the transition and specific testing conditions, these materials may exhibit either volatile or non-volatile resistive switching. Here, we focus on volatile threshold switching of these materials that are promising in demonstrating neuronal oscillator devices, although non-volatile multi-level resistive switching has also been reported when operating under different conditions (within hysteretic transition window) for both $VO_2$ and 1*T*-$TaS_2$.[54,136,147]

$VO_2$ crystals and thin films undergo a structural and insulator-to-metal transitions upon heating to about 340 K, exhibiting a resistivity drop of several orders of magnitude, as shown in Figure 3a. Resistive switching indued by applying voltage or current above a critical threshold, with negative differential resistance (NDR) commonly observed in current sweeping measurements, has been widely observed in both single crystals[148,149] and thin films[129] of $VO_2$, as illustrated Figure 7a. In 1980, Mansingh *et al*. systematically investigated the current-voltage (*I*-*V*) characteristics of a $VO_2$ single crystal at varying temperatures from 220 K to 325 K, while tracking the crystal temperature through a thermocouple attached to the sample surface.[32] They observed that lowering the ambient temperature increased the threshold voltage ($V_{th}$) upon switching but decreased the corresponding current ($I_{th}$) at $V_{th}$, resulting in a nearly constant threshold power ($V_{th} \times I_{th}$).[32] Notably, the measured crystal surface temperature rose only slightly above the ambient, remaining well below the bulk transition temperature of 340 K.[32] On this basis, they proposed a model involving a locally heated conducting channel of finite width, where switching is initiated once the channel temperature reaches the transition threshold of $VO_2$.[32] More recently, Kumar *et al*. utilized black-body emission mapping to spatially resolve the temperature distribution of a $VO_2$ thin film under varying applied currents, as shown in Figure 7b.[150] Similarly, Valle *et al*. employed spatiotemporal optical reflectivity measurements (Figure 7c) to characterize the growth and percolation dynamics of the metallic phase in $VO_2$, suggesting that the electrically driven insulator-to-metal transition can be explained just by considering the effect of Joule heating.[151] These results revealed the emergence and growth of a filament-shaped hot channel bridging the electrodes, further supporting the thermal nature of the switching.

Similar resistive switching behavior has also been reported in 1*T*-$TaS_2$. In 2015, Hollander *et al*. demonstrated electrically driven, reversible switching between the insulating CCDW phase and metallic NCCDW/ICCDW phases in mechanically exfoliated 1*T*-$TaS_2$ flakes at cryogenic temperatures, which was interpreted as a result of a carrier-assisted collapse of the Mott gap.[33] Later, Liu *et al*. exploited the unique transport properties of ultrathin 1*T*-$TaS_2$ (6 to 9 nm in thickness), where the low-temperature insulator-to-metal transition between the CCDW and NCCDW phases is suppressed, to demonstrate volatile resistive switching devices based on the NCCDW-to-ICCDW transition, which operated at a wide temperature range from 78 K to 320 K.[62] To evaluate the contributions of Joule heating to the resistive switching in 1*T*-$TaS_2$, Zhu *et al*. carried out temperature-dependent *I-V* measurements and estimated the local temperature rise due to Joule heating by establishing a simple analytical thermal model, as illustrated in Figure 7d.[63] The threshold Joule heating power required for switching reduces as the ambient temperature approaches $T_C$, consistent with a thermally driven picture.[63]

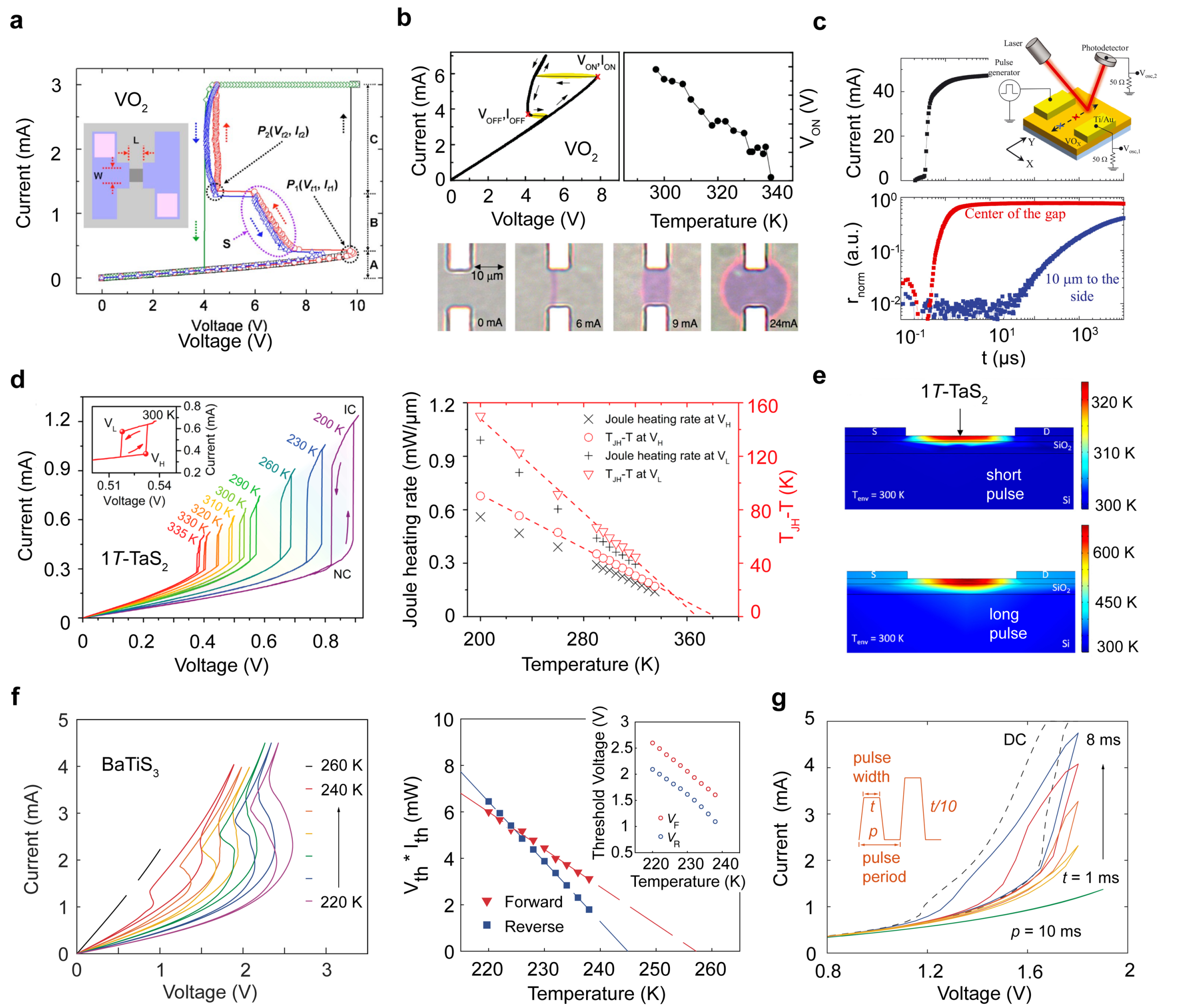

**Figure 7 Electrically driven volatile threshold switching and mechanisms.** (**a**) *I-V* characteristics of a $VO_2$ thin film devices with active channel dimensions of 10 μm × 10 μm in both *V*-mode (voltage sweeps) and *I*-mode (current sweeps). Figure adapted from Ref. [129] Copyright 2010, AIP Publishing. (**b**) Optical microscopic images (bottom) and electrical characteristics (top) of filaments observed between the electrodes at different applied currents. Figures adapted from Ref. [150] Copyright 2013, John Wiley & Sons. (**c**) Spatiotemporal characterization of the electrically induced insulator-to-metal transition in $VO_2$. Figures adapted from Ref. [151] Copyright 2021, American Association for the Advancement of Science. (**d**) Temperature dependent two-probe resistance of a 11.8-nm-thick 1*T*-$TaS_2$ flake (left) and the associated Joule heating rate analysis. Figures adapted from Ref. [63] Copyright 2018, ACS Publications. (**e**) Cross-sectional view of the simulated temperature distribution near a 1*T*-$TaS_2$ device channel, for applied current pulses of short (8 ns) and long (13.3 μs) durations. Figures adapted from Ref. [152] Copyright 2021, AIP Publication. (**f**) Temperature dependent four-probe resistance of a bulk $BaTiS_3$ device from 220 K to 240 K (left) and the associated thermal power analysis (right). (**g**) Pulsed *I-V* characterization of $BaTiS_3$. (**f**) and (**g**) Figures adapted from Ref. [34] Copyright 2023, John Wiley & Sons.

Pulsed *I-V* measurements also provide insights into the switching mechanism by deconvoluting the thermal and electric-field contributions that can otherwise be conflated in conventional DC measurements. In a pulsed *I-V* measurement, an arbitrary waveform generator produces a single or a series of pre-programmed voltage/current pulses and sends them to the device under test, while the response is recorded by an oscilloscope or other high-speed measurement instrument. effects.[153] In 2021, Mohammadzadeh *et al*. studied the NCCDW-to-ICCDW switching in exfoliated 1*T*-$TaS_2$ flakes using electrical pulses ranging from 8 ns to ~ 13 μs.[152] They found that the reconstructed pulsed *I-V* characteristics exhibited no hysteresis for pulse durations below ~ 200 ns, whereas longer pulses induced pronounced hysteresis that peaks at some particular value of pulse duration.[152] COMSOL-based simulations show that the heating is insufficient to drive the NCCDW-to-ICCDW transition for 8-ns pulse duration (Figure 7e, top), while the local temperature of the 1*T*-$TaS_2$ channel rises well above 350 K with a 13.3 μs duration (Figure 7e, bottom).[152] Mohammadzadeh *et al*. further suggested that, despite the thermal origins of the switching, GHz-level switching speed are still feasible in thin 1*T*- $TaS_2$ devices by optimizing device geometry and thermal resistance.[152]

First observation of electrically driven reversible resistive switching in $BaTiS_3$ was reported in 2023, shortly after the initial discovery of phase transitions in this material.[34] $BaTiS_3$ exhibits a transition from a semiconducting room-temperature phase to a CDW state near 250 K, featuring a resistance jump of two- to threefold and a thermal hysteresis of ~ 10 K,[56] as illustrated in Figure 3e. Electrically, $BaTiS_3$ exhibits a threshold switching behavior below its transition temperature $T_c$: the *I-V* characteristics show an abrupt switching to a conductive state above a threshold voltage ($V_F$) during forward scan, and a return to its original high-resistance state below another critical voltage ($V_R$) during reverse scan, forming a characteristic hysteresis window,[34] similar to those reported for both $VO_2$ and 1*T*-$TaS_2$. When driving the device with a current source, an "S-type" NDR regime is also observed.[34]

To assess the contribution of Joule heating in $BaTiS_3$ resistive switching, Chen *et al*. carried out thermal analysis based on four-probe *I-V* sweeps at various temperatures across the CDW transition, as shown in Figure 7f.[34] They noted that the critical voltage ($V_{th}$) required to switch the device increases with decreasing temperature, as does the threshold current ($I_{th}$).[34] The threshold thermal power ($P_{th}$) generated by Joule heating was then estimated using $P_{th} = V_{th} \times I_{th}$ for both forward and reverse scans at each temperature. Unlike $VO_2$, where $P_{th}$ remains largely temperature-independent and hence the switching is achieved through a conducting channel (or filament),[32] $P_{th}$ in $BaTiS_3$ was found to decrease linearly as the temperature approached $T_c$. Two characteristic temperatures of 245 K and 258 K were extrapolated at which the threshold thermal power equals to zero, which aligns with the transition temperatures extracted from low-field transport measurements.[34] Moreover, pulsed *I-V* measurements were carried out at 210 K with varying pulse widths ranging from 1 ms to 8 ms and a constant pulse period of 10 ms.[34] The reconstructed I-V curves with voltages swept between 0.8 and 1.8 V are displayed in Figure 7g. Decreasing the pulse width from 8 ms to 1 ms, while maintaining the same pulse period and voltage sweep ranges, reduced the contribution of Joule heating and in turn narrowed the hysteresis window and raised the threshold switching voltage.[34] Notably, no hysteresis was observed at 1 ms pulse width. Therefore, Chen *et al*. suggested that the switching in $BaTiS_3$ is primarily thermally driven, as reducing pulse width decreases Joule heating power without altering the applied electric field.[34]

## 7. Self-sustained voltage oscillations

This section surveys the development of the self-sustained phase-change oscillator devices of $VO_2$, 1*T*-$TaS_2$, and $BaTiS_3$, by leveraging the electrically driven volatile resistive switching in the systems. The detailed device geometries, the associated threshold switching behavior, and voltage oscillation performance are listed in Table 3.

Self-sustained oscillations in $VO_2$ were first reported in 1975 by Taketa *et al*., who employed a bulk crystal device (~ mm channel length) connected in series with a load resistor under an applied DC voltage.[149] The oscillation frequency at room temperature was approximately 5 kHz, and it decreases as the ambient temperature approached the transition temperature.[149] In 2008, Sakai *et al*. achieved a substantially higher oscillation frequency of ~ 550 kHz in a $VO_2$ thin-film device (10 µm × 10 µm × 0.22 µm), fabricated by PLD and subsequent dry etching.[126] As shown in Figure 8a and 8b, similar device performance with oscillation frequency > 0.5 MHz was demonstrated by Kim and Lee *et al*. on sol-gel prepared $VO_2$ thin-film devices with comparable lateral sizes (10 µm × 10 µm × 0.1 µm).[129,154] In 2011, Seo *et al*. experimentally investigated dimensional effects on $VO_2$ oscillators in planar device geometries, demonstrating a decrease in oscillation frequency as either the channel length or width increased.[130] This finding suggests that scaling the active device region could lead to higher oscillation frequencies, pending other limiting factors such as contact resistances or intrinsic capacitance.

Alternatively, in 2014, Beaumont *et al*. investigated out-of-plane $VO_2$ devices (3 μm × 3 μm) integrated in crossbars geometry and realized oscillation frequencies of up to 300 kHz.[155] They noted that such metal-$VO_2$-metal configurations not only minimize the device footprint to enhance circuit integration, but also reduce the threshold voltage for triggering insulator-to-metal transition by shortening the conduction path to the film thickness (e.g., <100 nm).[155] However, using regular Ti/Au/Ti bottom electrodes largely constrained the achievable $VO_2$ film quality, likely due to limited growth temperatures or poor interfaces, resulting in low resistance jump across the transition.[155] Similar limitations have been reported using $SrRuO_3$ or Pt bottom electrodes, where resistance changes across the transition are often below one order of magnitude.[132] In 2015, Mian *et al*. addresses these challenges by incorporating TiN as the bottom electrode, improving the interface and overall $VO_2$ film quality in an out-of-plane device configuration.[131] This approach yielded self-oscillations of up to 9 MHz with low threshold switching voltages and currents.[131] Further optimization of thin-film deposition conditions and electrode-integration strategies, aimed at minimizing contact resistances and active channel sizes, could further increase the oscillation frequencies of $VO_2$ devices, thereby making them more attractive for practical high-frequency applications.

**Table 3 Performance of phase-change volatile switching and oscillator devices of $VO_2$, 1*T*-$TaS_2$, and $BaTiS_3$.**

| **Material** | **Device Geometry** | | | **Temp.** | **Threshold Switching** | | **Oscillator Device** | | | **Ref.** |
|---|---|---|---|---|---|---|---|---|---|---|
| | Device Config. | Electrode | Channel Length | | $V_{th}$ | $I_{th}$ | $V_S$ or $I_S$ | $f$ | Year | |
| $VO_2$ | Planar crystal | N/A | 2 mm | RT | 16 V | N/A | N/A | 5 kHz | 1975 | 149 |
| | Planar film | Ti/Au | 10 μm | RT | 7 V | ~ 1 mA | ~10 V | >550 kHz | 2008 | 126 |
| | Planar film | Ti/Au | 5 μm | RT | 3 V | 0.5 mA | 10 V | 0.8 MHz | 2011 | 130 |
| | Vertical film | Ti/Au | 130 nm | RT | 0.8 V | 0.6 mA | 1 mA | 300 kHz | 2014 | 155 |
| | Vertical film | TiN | 200 nm | RT | 1.8 V | 80 μA | 13 V | 9 MHz | 2015 | 131 |
| 1*T*-$TaS_2$ | Planar flake | Pd/Au | 0.5-1 μm | RT | ~0.8 V | 3 mA | ~ 4 V | 2 MHz | 2016 | 62 |
| | Planar flake | Cr/Au | 0.7 μm | RT | 0.78 V | 2.8 mA | ~ 3.9 V | 1.1 MHz<br>0.8 MHz w. light | 2018 | 63 |
| $BaTiS_3$ | Planar crystal | Ti/Au | 80 μm | 220 K | 7 V | ~ 1 mA | ~ 20 V | 16 Hz | 2023 | 34 |
| | | | 10 μm | 130 K | N/A | N/A | 1.23 mA | 0.9 kHz | | |

Despite extensive research on bulk 1*T*-$TaS_2$ crystals in the 1970s and 1980s, including several studies on CDW-associated metal-insulator transitions,[52,84] the phase-change oscillator applications based on this material were not realized until the mid-2010s.[62,63] Much of the current electrical studies of 1*T*-$TaS_2$ relies on mechanically exfoliated micro-scaled flakes, whose lateral dimensions and production yield remain limited. In 2016, Liu *et al*. constructed the first room-temperature CDW oscillator using thin 1*T*-$TaS_2$ flakes.[62] Their device exploits the electrically driven NCCDW-to-ICCDW transition near 350 K, achieving a maximum oscillation frequency of ~2 MHz in a simple oscillation circuit composed of a two-terminal 1*T*-$TaS_2$ device in series with a load resistor,[62] as shown in Figure 8c and 8d. They further integrated a top-gated graphene field-effect transistor (FET) with the 1*T*-$TaS_2$ device, enabling tunable oscillation frequencies through gate-mediated adjustments of the load resistance.[62] Although the 1*T*-$TaS_2$ flakes used for demonstration were only 6–9 nm thick, the same room-temperature oscillation behavior could, in principle, be observed in thicker 1*T*-$TaS_2$ crystals, provided that the NCCDW-to-ICCDW transition persists. Suppressing the low-temperature CCDW-to-NCCDW transition in such thin 1*T*-$TaS_2$ extends the oscillator's operation window to much lower temperature ranges without interference from the low-temperature transition. This disappearance of the CCDW-to-NCCDW transition in sufficiently thin 1*T*-$TaS_2$ flakes is typically attributed to extrinsic surface oxidation effects, as discussed by Tsen *et al*. in 2015.[137] Furthermore, Zhu *et al*. developed light-tunable CDW oscillators based on 1*T*-$TaS_2$, whose oscillation frequency can be well modulated by varying the illumination intensity through the laser thermal effect.[63]

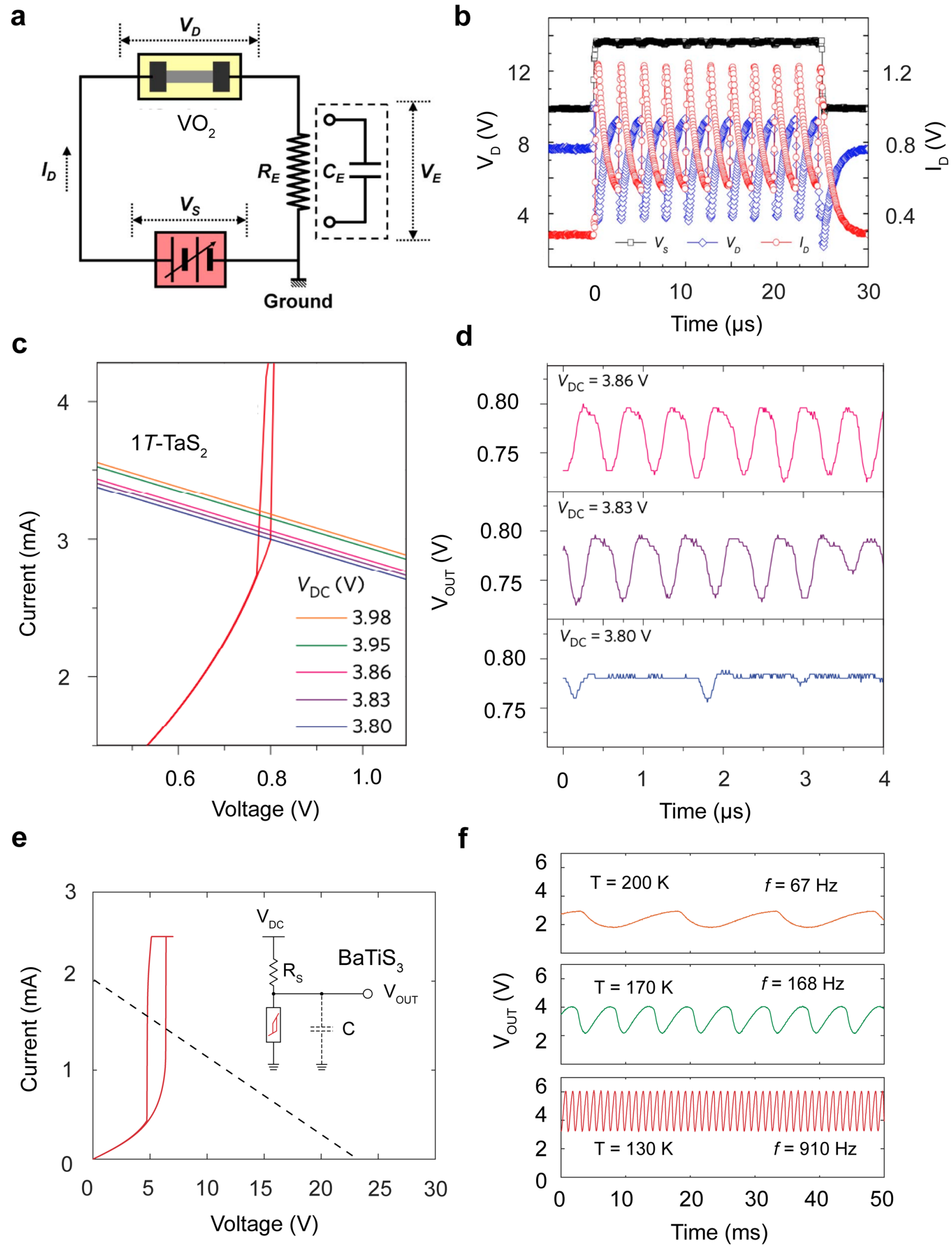

**Figure 8 Self-sustained voltage oscillations.** (**a**) Schematic diagram of the electrical circuit used for the generation of the MIT oscillation in a $VO_2$ thin film device. (**b**) Waveform of $V_D$ and $I_D$ of the device measured at room temperature. (**a**) and (**b**) Figure adapted from Ref. [129] Copyright 2010, AIP Publishing. (**c**) *I-V* characteristics of a 1*T*-$TaS_2$ flake device at room temperature and load lines of the series resistor at different $V_{DC}$ values. The oscillator circuit starts to oscillate across the threshold condition of $V_{DC}$ = 3.8 V. (**d**) Voltage oscillations of a 1*T*-$TaS_2$ flake device under different $V_{DC}$ values. (**c**) and (**d**) Figures adapted from Ref. [62] Copyright 2016, Springer Nature. (**e**) *I-V* characteristics of a two-terminal $BaTiS_3$ device at 220 K. The dashed line represents a series resistor load line for stable voltage oscillations and the inset shows a schematic illustration of the oscillator measurement circuit. (**f**) Oscillation waveforms of a $BaTiS_3$ device operated at different temperatures. (**e**) and (**f**) Figures adapted from Ref. [34] Copyright 2023, John Wiley & Sons.

As discussed earlier, phase-change oscillators have been experimentally demonstrated in both $VO_2$ and 1*T*-$TaS_2$ with oscillation frequencies reaching the MHz regime, although the underlying switching mechanisms may differ. $BaTiS_3$ undergoes a unique semiconductor-to-CDW transition near 250 K, resulting in a two- to threefold resistivity jump and a thermal hysteresis window of ~ 10 K.[34,56] The overall shape of this transition, in terms of temperature-dependent resistivity, resembles that of the MIT in $VO_2$ or the ICCDW-to-NCCDW transition in 1*T*-$TaS_2$, suggesting the feasibility of inducing voltage oscillations in $BaTiS_3$. In 2023, Chen *et al*. reported the first phase-change oscillator based on single-crystal $BaTiS_3$ operating at 220 K (~ 30 K below its transition temperature), achieving an oscillation frequency of ~16 Hz by connecting a two-terminal $BaTiS_3$ device in series with a load resistor ($R_S$) under a DC bias,[34] as shown in Figure 8e. The oscillation arises from volatile threshold resistive switching, a common feature of all three phase-change materials discussed in this review, and reflects the periodic changes of electrical resistance triggered by repetitive local heating and cooling cycles across the phase transition boundary.[34] When the voltage applied across the $BaTiS_3$ channel exceeds the critical voltage ($V_F$), a transition to the low-resistance state is initiated due to Joule heating ($P = \frac{{V_{DC}}^2}{(R+R_S)^2/R}$), resulting in a sudden increase in current and a subsequent increase in voltage across the load resistor[34]. The Joule heating powder ($P$) is insufficient to maintain the temperature in the low-resistance regime, because $P$ decrease as $R$ is reduced (for $R < R_S$), driving $BaTiS_3$ back to the CDW state. The cycle repeats and yields self-sustained voltage oscillations.[34]

Despite this successful demonstration in a relatively new phase-change material, the oscillation frequencies of $BaTiS_3$ single-crystal devices remain orders of magnitude lower than those reported for $VO_2$ and 1*T*-$TaS_2$, due to combined effects of non-ideal sample morphologies, device geometries, and thermal managements.[34] Inspired by earlier efforts with $VO_2$ single-crystal oscillators,[149] Chen *et al*. employed a thermal management strategy to enhance the oscillation frequencies of $BaTiS_3$ devices.[34] Figure 8f shows that the oscillation frequency of the same device increased from 67 Hz to 910 Hz by lowering the operating temperature from 200 K to 130 K, benefited from an enhanced cooling efficiency.[34] However, maintaining stable CDW oscillations below ~ 130 K proved challenging, as the low-temperature structural transition in $BaTiS_3$ starts to interfere with the switching behavior.[56] Strategies that can suppress this low-temperature transition could potentially improve the device performance but require further research investigations. Additionally, Chen *et al*. improved oscillation frequency by reducing the channel length, a method previously validated in other oscillating systems such as $VO_2$.[130] Both operating at 170 K, a $BaTiS_3$ device with a 5 μm channel outperformed a 10 μm device by over a factor of three in oscillation frequency.[34] This effect likely stems from more efficient local heating and cooling in smaller channels. Further reduction of the $BaTiS_3$ channel dimensions, including thickness, is expected to oscillation frequency significantly. Consequently, synthesizing high-quality $BaTiS_3$ thin film with intrinsic phase-change properties would be highly desired for their practical electronic device applications, benefiting from the precise thickness control and the fabrication scalability.

## 8. Device implementation of phase-change neuronal oscillators

As for the device implementation aspect, phase-change neuronal oscillators can be used individually or integrated into high-level circuits to not only emulate both deterministic and stochastic dynamics of biological neurons, but also leverage the dynamics to solve various NP-hard optimization tasks accurately and efficiently that are otherwise challenging using conventional von Neumann processors. In this section, we discuss potential neuronal device applications of these phase-change materials at the circuit integration level. Note that due to distinctive differences in their technological readiness and literature availability, as illustrated in Table 4, the majority of the device application examples covered in this section are based on $VO_2$ oscillators.

Electronic devices based on threshold resistive switching in phase-change materials are promising candidates for emulating biological-neuron-like temporal dynamics at different levels of complexities. Based on the classification criterions proposed by Kumar *et al*., most selector-type devices exhibiting volatile switching, either voltage controlled or current controlled, are categorized as first-order, whose dynamical complexity arises from the changing of internal temperatures during the switching.[156] This class includes the two-terminal $VO_2$, 1*T*-$TaS_2$, and $BaTiS_3$ devices discussed in Section 6, even though the underlying operating mechanisms may vary. Second-order neuronal complexity such as self-sustained oscillations can be realized by integrating a first-order device into a relaxation oscillator circuit, as shown in Section 7.[156] The higher-order complexity is achieved *via* two alternating dynamical processes: i.e., the charging-discharging dynamics of the capacitor and the internal thermodynamics of volatile switching.[156] Standalone second-order devices are also possible when their built-in capacitance is sufficiently large to sustain the oscillations without the need for an external parallel capacitor, as demonstrated with $NbO_2$.[157]

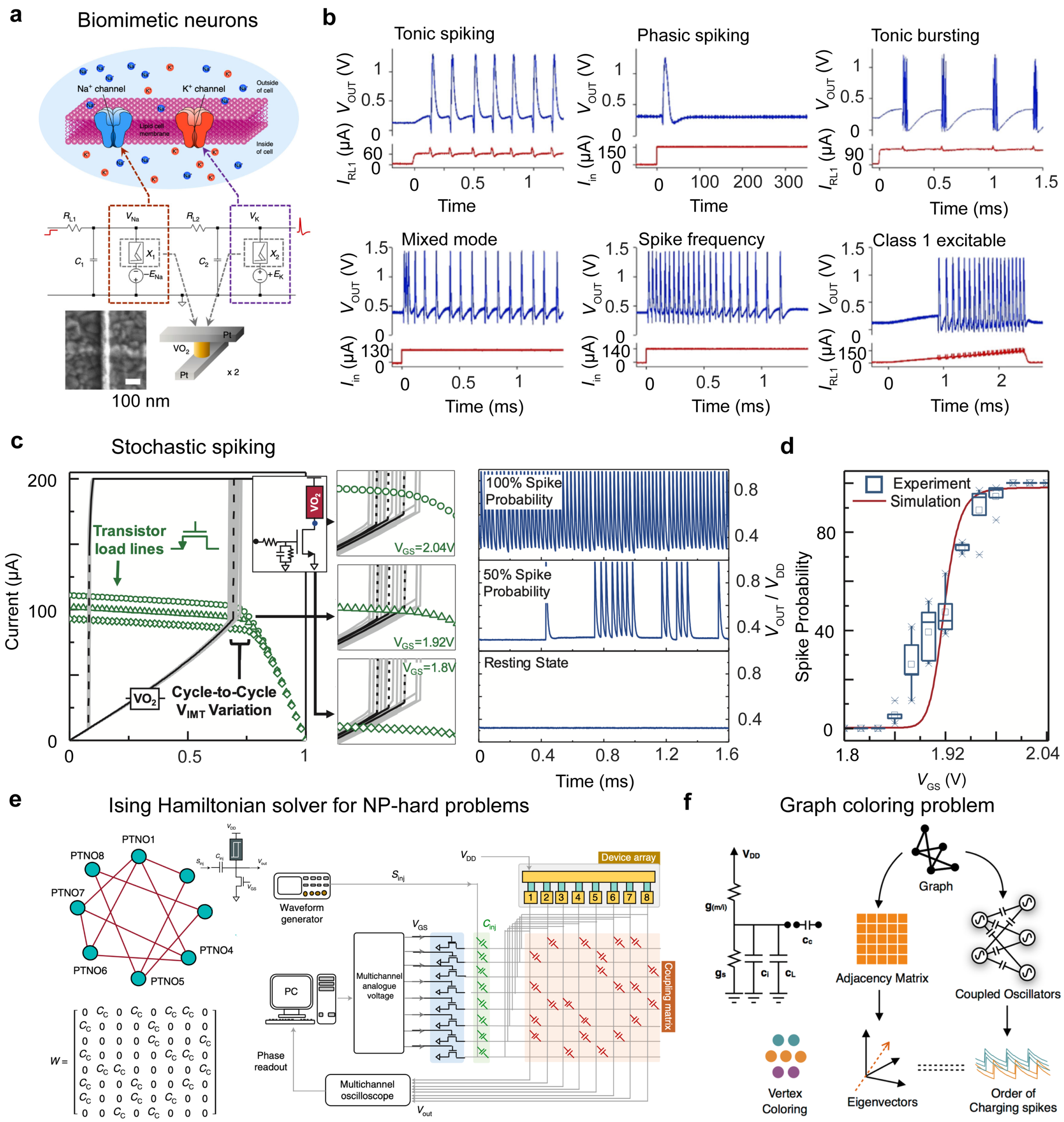

**Figure 9 Device implementation of phase-change neuronal oscillators.** (**a**) Circuit diagrams of a two-channel artificial neuron to emulate the biological neuronal dynamics. (**b**) Six representative biological neuron spiking behaviors experimentally demonstrated in $VO_2$ neurons. (**a**) and (**b**) Figures adapted from Ref. [158] Copyright 2018, Springer Nature. (**c**) Stochastic spiking of a $VO_2$ oscillator due to cycle-to-cycle variations in threshold switching voltages. (**d**) The associated spike probability that follows a sigmoidal pattern. (**c**) and (**d**) Figures adapted from Ref. [159] Copyright 2017, IEEE. (**e**) Schematic illustration of an Ising Hamiltonian solver for combinatorial optimization problems using coupled stochastic $VO_2$ oscillators. Figures adapted from Ref. [160] Copyright 2021, Springer Nature. (**f**) Schematic illustration of the process flow to solve the graph coloring problem using coupled $VO_2$ oscillatory network. Figures adapted from Ref. [161] Copyright 2017, Springer Nature.

Achieving even higher-order complexities often requires multiple coupled circuit elements. For instance, Yi *et al*. reported a non-transistor fourth-order neuronal system employing two volatile $VO_2$ switching devices and a series of circuit elements to emulate the biologically accurate Hodgkin-Huxley neuron model.[158] This approach successfully reproduced all 23 neuronal functionalities, including spiking, bursting and mixed mode firing. Figure 9a illustrates the structure of a biological neuron membrane, highlighting $Na^+$ and $K^+$ channels. In the corresponding circuit diagram, two $VO_2$ neuronal devices ($X_1$ and $X_2$) are coupled with two capacitors ($C_1$ and $C_2$) and two series load resistors ($R_{L1}$ and $R_{L2}$). When driven by a steady DC current, this circuit can produce periodic bursts of spikes, a hallmark of tonically active neurons.[158] By either replacing the load resistor $R_{L1}$ with a capacitor $C_{in}$, or by inserting $C_{in}$ before $R_{L1}$ in the tonic neuron circuit, one can emulate another basic type of neurons, i.e., phasically active neurons, characterized by a single spike fired upon the onset of a DC input followed by quiescence afterwards.[158] Moreover, by placing a capacitor $C_{in}$ in parallel with $R_{L1}$, a mixed-mode neuron can be simulated, which features a phasic burst followed by a series of tonic spiking upon applying the DC input. Figure 9b shows part of the 23 experimentally demonstrated biological neuron spiking behaviors. A detailed circuit design and neuron emulation results can be found in the original reference.[158]

The stochastic dynamics represents another important aspect of neuronal complexities that plays key roles in signal and information processing.[162,163] Recently, several phase-change materials,[159,164,165] including $VO_2$ and 1$T$-$TaS_2$, have been explored to demonstrate stochastic artificial neurons. For instance, $VO_2$-based devices show cycle-to-cycle variations in the threshold switching voltages, as illustrated in Figure 3c (left).[159] Therefore, when the switching devices are integrated to construct neuronal oscillators, depending on where the electrical load line locates (controlled by $V_{GS}$), the status of output waveforms vary (Figure 9c, right).[159] The neuron sits in a resting state when the load line crosses the stable high-resistance state (e.g., $V_{GS}$ = 1.8 V). As $V_{GS}$ increases (e.g., $V_{GS}$ = 1.92 V), the load line occasionally crosses both boundaries of hysteresis window (dashed lines) due to cycle-to-cycle $V_{th}$ variation, leading to probabilistic firing of oscillations. The spike probability follows a sigmoidal pattern, as shown in Figure 9d. Further increases in $V_{GS}$ (e.g., $V_{GS}$ = 2.04 V) give rise to varying amplitude of the oscillations (determined by the two threshold voltages), although the spike probability reaches 100%.[159] Similar idea has been applied to 1*T*-$TaS_2$ stochastic neuronal oscillators later by Liu *et al*. to study detailed statistical characteristics of these artificial neurons.[165]

One of the most promising use cases of phase-change neuronal oscillators is to effectively solve complex computational problems such as large-scale combinatorial optimization. These problems belong to a so-called non-deterministic polynomial time (NP)-hard complexity class and are important in various practical applications such as machine learning, route planning, and resource allocation, however, the computational resources required to find an optimal solution to them using conventional von Neumann computers increases rapidly with problem size, making them almost impractical to solve effectively and energy efficiently.[166] In 2014, Lucas proposed that many NP-hard optimization problems can be mapped into the problem of finding the ground state of an Ising model with the Ising Hamiltonian $H$ (consisting of Ising spin $\sigma$ and the coupling matrix $J$), offering a potential physical solution for such problems.[167]

In 2021, Dutta *et al*. constructed an Ising Hamiltonian solver based on a network of eight electrically coupled stochastic $VO_2$ phase-transition nano-oscillators (PTNOs), as illustrated in Figure 9e.[160] The artificial Ising spin $\sigma$ is emulated by the binary degree of freedom in phases due to the second-harmonic injection-locking (SHIL) phenomenon, while the coupling matrix $J$ is emulated by the capacitive coupling between $VO_2$ neuronal oscillators with the matrix $W$. When the oscillator is perturbated with the injection-locking signal ($S_{inj}$) at the second harmonic ($f_{inj} \sim 2f_0$), the output waveform shows both in-phase (40˚, representing up-spin) and out-of-phase (220˚, representing down-spin) injection-locking configuration, while the first-harmonic perturbation only gives rise to a constant 80˚ phase-locking configuration.[160] Dutta *et al*. further implemented this oscillator network to experimentally solve an NP-hard MaxCut graph problem by minimizing the Ising Hamiltonian *H*. Note that this platform features an exceptionally high energy efficiency of $1.3 \times 10^7$ solutions per second per watt for 100-node problems, offering several orders of magnitude improvement over conventional approaches using CPUs or GPUs.[160] Figure 9f illustrates another example of solving an NP-hard problem (vertex graph coloring problem) using $VO_2$ oscillators capacitively coupled in a network same as the input graph. The experimentally obtained order of oscillator phases is then correlated to the eigenvectors of the adjacency matrix of the input graph, which in turn are related to the solution of the graph coloring problem.[161]

It is worth noting that the device implementation examples discussed in this section are not exclusively limited to $VO_2$-based devices. One can certainly extend these applications to construct 1*T*-$TaS_2$ and $BaTiS_3$-based neuronal oscillatory circuits for biomimetic emulation and solving optimization problems, considering their similarities in demonstrated volatile threshold switching and oscillatory behaviors. However, unlike $VO_2$, whose high-quality thin-film synthesis and large-scale device integration are relatively well established, the scalable fabrication of 1*T*-$TaS_2$ microflakes and $BaTiS_3$ bulk crystals remains challenging. Therefore, advancing the synthesis of 1*T*-$TaS_2$ and $BaTiS_3$ thin films with pristine CDW phase transitions is essential to unlock energy-efficient, chalcogenide-based neuronal device applications in the future.

## 9. Summary and outlook

In this review, we have surveyed recent research progresses on three representative phase-change materials exhibiting metal-to-insulator transitions – specifically, the correlated oxide $VO_2$, the transition metal dichalcogenide 1*T*-$TaS_2$, and the hexagonal chalcogenide perovskite $BaTiS_3$ – and their potential applications in neuronal devices. Despite differences in chemical composition, morphology, underlying phase-change mechanisms, and levels of technological readiness, these three materials display similar volatile threshold resistive switching across the associated transitions and exhibit self-sustained voltage oscillations, making them promising candidates for neuronal device applications. From the perspective of material development and device implementation, we sequentially discussed their coincident electronic and structural phase transitions, material synthesis, device fabrication, transport characterization, resistive switching, sustained oscillation, and neuronal device integration.

Table 4 summarizes an overall assessment of the technological readiness of $VO_2$, 1*T*-$TaS_2$, and $BaTiS_3$ for practical electronic device applications, in terms of material development, device processing, operation, and integration. Despite these advances, there are still significant challenges for practical implementation of these volatile switching-based phase-change neurons in neuromorphic computing. For instance, to be compatible with the processing conditions for fabricating artificial synapses or CMOS-based chips, the material deposition temperature must be maintained below 400˚C, or alternative transfer-based heterogeneous integration approaches must be developed. Among these three materials, $VO_2$ is currently the only system demonstrating high-quality thin-film deposition processes that produce an metal-to-insulator transition comparable to that of a bulk crystals.[96] Nonetheless, depositing at such reduced temperatures on non-ideal substrates may compromise materials quality, leading to changes of transport behaviors such as a shift of transition temperature and a decrease in the amplitude of the resistance jump.[168]

**Table 4 Technological readiness of $VO_2$, 1*T*-$TaS_2$, and $BaTiS_3$ for practical electronic device applications.**

| **Material** | **Form** | **Device Processing** | | **Device Operation** | | | **Integration** | **Tech. Readiness** |
|---|---|---|---|---|---|---|---|---|
| | | Fabrication Method | Air / thermal stability | Temp. | Best Performance | Channel Size | CMOS Compatibility | |
| $VO_2$ | Thin film | lithography | Excellent | RT | 9 MHz | 200 nm (film thickness) | High | High |
| 1*T*-$TaS_2$ | Cleaved flakes | Mechanical exfoliation + ebeam lithography | Poor at thin limit | RT | 2 MHz | 0.5-1 µm | Medium | Low |
| $BaTiS_3$ | Bulk crystals | Polymeric planarization + lithography | Good | Cryogenic | ~ 0.9 kHz | 10 µm | Low | Very low |

Even more research efforts are required for the two CDW-based materials, i.e., 1*T*-$TaS_2$ and $BaTiS_3$. So far, single crystals of 1*T*-$TaS_2$ and $BaTiS_3$ are still being used to examine their intrinsic material properties and demonstrate proof-of-concept prototype devices. Although several back-end-of-line (BEOL)-compatible 2D and 3D integration approaches have been recently reported,[169-171] such labor-intense integration procedures and limited device packing density would inevitably increase the overall fabrication cost when dealing with 1*T*-$TaS_2$ and $BaTiS_3$. Developing high-quality thin-film synthesis approaches for these materials is itself a demanding but rather challenging task, considering the susceptibility of external factors such as strain, defects, and doping on CDW phase transitions.[56,83,172] Alternatively, one may also leverage this sensitivity to intentionally tune the phase transition temperatures as needed and to maximize the overall device performance.

# Author declarations

## Acknowledgements

We gratefully acknowledge support from an ARO MURI program (W911NF-21-1-0327), an ARO grant (W911NF-19-1-0137) and National Science Foundation (DMR-2122071).

## Conflict of interest

The authors declare no competing financial interests.

# Biographies

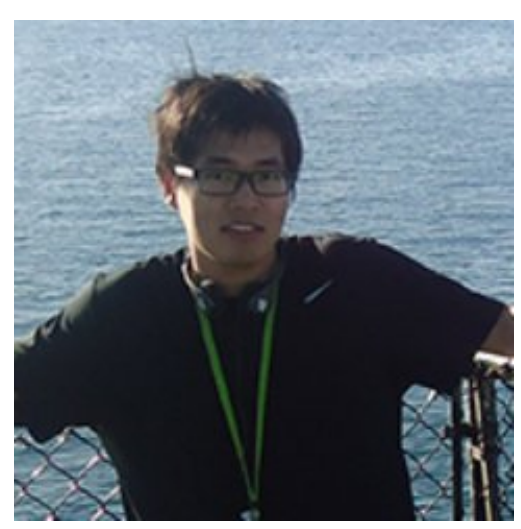

**Huandong Chen** is currently a postdoctoral researcher in Condensed Mater Physics and Materials Science Department at Brookhaven National Laboratory, studying low-dimensional quantum materials and devices using Scanning Tunneling Microscopy (STM). He received his Ph.D. degree in Materials Science from University of Southern California in 2023, where he has worked on bulk single crystal synthesis, transport studies and electronic and optoelectronic devices of novel complex chalcogenides.

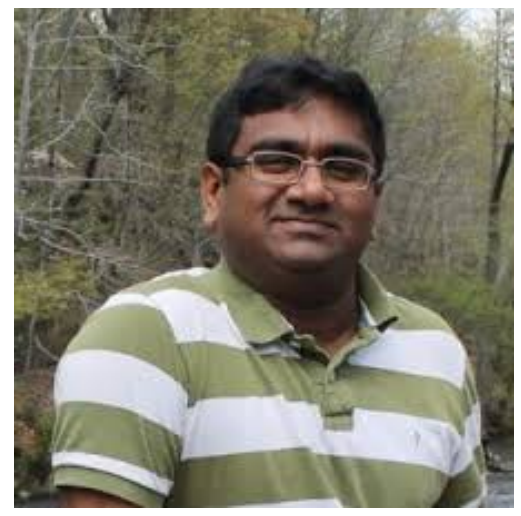

**Jayakanth Ravichandran** is currently a Professor of Chemical Engineering and Materials Science and Electrical and Computer Engineering at University of Southern California, and the holder of the Philip and Cayley MacDonald Endowed Early Career Chair. He is also a co-director of the Core Center for Excellence in Nano Imaging. He completed his graduate work at University of California, Berkley in Applied Science and Technology, and postdoc in Physics at Columbia University and Harvard University. His research interests are in the broad area of electronic and photonic materials and devices.

# TOC

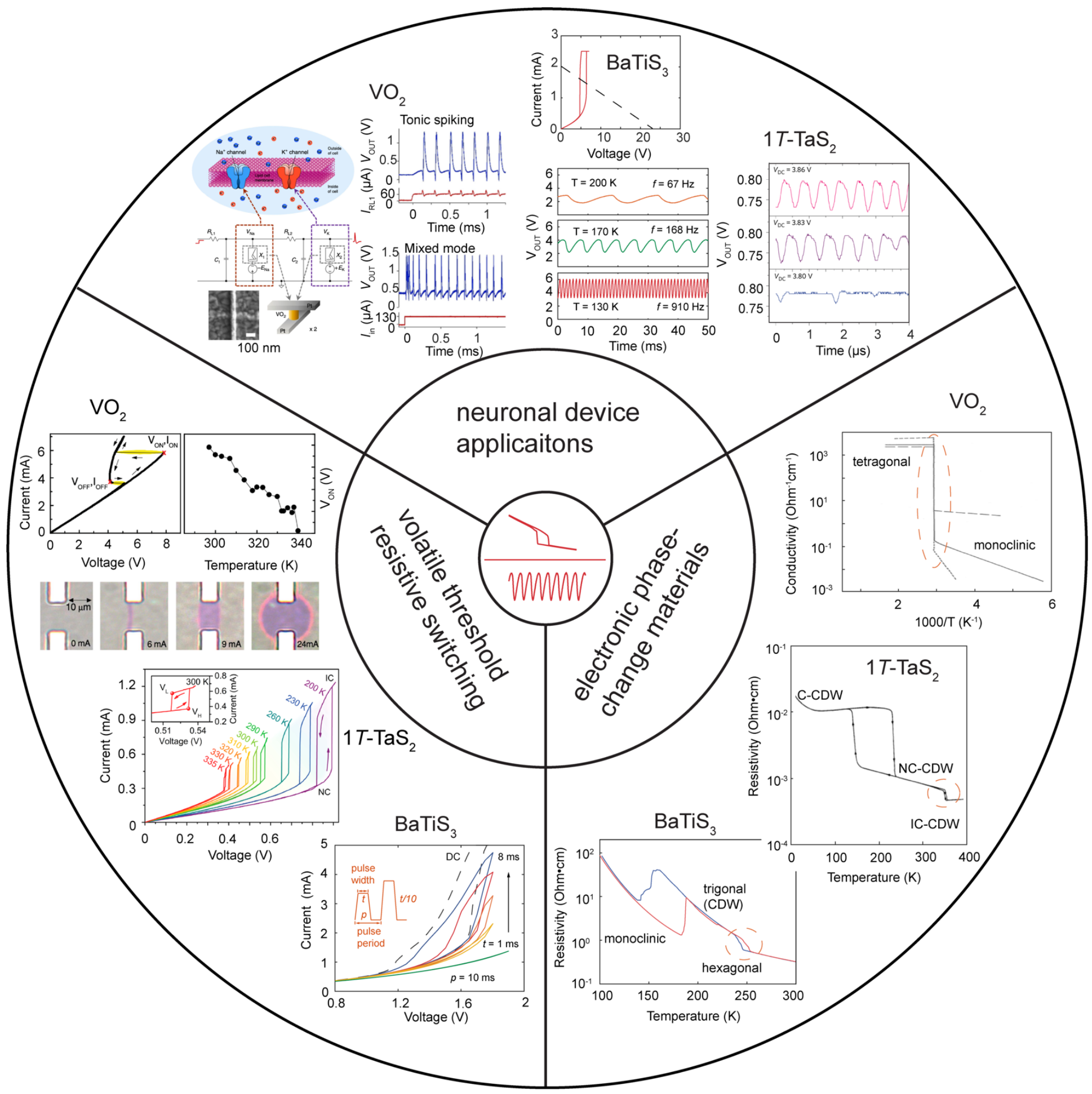

neuronal device applicaitons
volatile threshold resistive switching
electronic phase-change materials
VO2
Tonic spiking
Mixed mode
BaTiS3
1T-TaS2
tetragonal
monoclinic
C-CDW
NC-CDW
IC-CDW
trigonal (CDW)
hexagonal
100 nm